\documentclass[11pt]{article}

\usepackage{graphicx,epsf, epsfig, amssymb, amsmath, amsfonts, mathrsfs}
\usepackage{longtable}
\usepackage{bm}
\usepackage{color}
\usepackage{enumitem}
\usepackage{float}
\usepackage{caption}
\usepackage{subcaption}
\usepackage{upgreek}
\usepackage{cite}
\usepackage{mathtools}
\usepackage{titlesec}
\usepackage{lipsum}
\usepackage{xcolor}
\usepackage[section]{placeins}
\usepackage[colorlinks=true,linktocpage=true,linkcolor=blue,citecolor=blue]{hyperref}
\usepackage{tensor}
\usepackage{multirow}
\usepackage[utf8]{inputenc}
\usepackage{authblk}
\usepackage{array}
\usepackage{breqn}

\newcommand{\be}{\begin{equation}}
\newcommand{\ee}{\end{equation}}
\newcommand{\bea}{\begin{eqnarray}}
\newcommand{\eea}{\end{eqnarray}}

\newcommand{\eqn}[1]{(\ref{#1})}

\newcommand{\dmoff}[1]{}

\newcommand{\pa}{\partial}

\newcommand\blfootnote[1]{%
  \begingroup
  \renewcommand\thefootnote{}\footnote{#1}%
  \addtocounter{footnote}{-1}%
  \endgroup
}
\usepackage{comment}
\newcommand{\customrule}{\vspace{0.5cm}\noindent\textcolor{gray}{\rule{\textwidth}{0.5pt}}\vspace{0.5cm}}

\usepackage
[
a4paper,% other options: a3paper, a5paper, etc
left=1.5cm,
right=1.5cm,
top=3cm,
bottom=4cm,
]{geometry}

\numberwithin{equation}{section} %For numbering equations according to section

\titleclass{\subsubsubsection}{straight}[\subsubsection]
\newcounter{subsubsubsection}[subsubsection]
\renewcommand\thesubsubsubsection{\thesubsubsection.\arabic{subsubsubsection}}

\titleformat{\subsubsubsection}[block]
  {\normalfont\normalsize\bfseries}{\thesubsubsubsection}{1em}{}
\titlespacing*{\subsubsubsection}
  {0pt}{3.25ex plus 1ex minus .2ex}{1em}

\makeatletter
\def\toclevel@subsubsubsection{4}
\def\l@subsubsubsection{\@dottedtocline{4}{7em}{4em}}
\makeatother

\author{Etevaldo dos Santos Costa Filho}

\affil[]{Center for Research and Development in Mathematics and Applications
(CIDMA), Department of Mathematics, University of Aveiro, 3810-193 Aveiro,
Portugal}

\begin{document}
	\title{\bf From Rotating Attractors to Extremal Black Holes \\with Axionic Hair}
	\date{May 2026}

	\maketitle

	\begin{abstract}
We study extremal, rotating black holes in four-dimensional Einstein-Maxwell-axion (EMA) theory through a combined near-horizon and bulk analysis. At the level of the near-horizon extremal geometry (NHEG), using the entropy-function formalism, we prove that regular rotating attractors with axionic hair exist only for configurations that are purely electrically or purely magnetically charged; regular rotating dyonic attractors are excluded by the axion equation of motion, a result that we established perturbatively and non‑perturbatively within the NHEG system.  On the global side, we construct families of asymptotically flat, rotating extremal EMA black holes that interpolate to the electric NHEG branch, confirming that horizon data are fixed by extremization of the entropy-function and decoupled from asymptotic moduli in line with the attractor mechanism. 
	\end{abstract}
\vfill
\blfootnote{E-mail: {\tt etevaldo.s.costa@ua.pt}}

	\newpage
	\customrule
	\tableofcontents

\customrule

	%%%%%%%%%%%%%%%%%%%%%%%%%%%%%%%%%%%%%%%%%%%%%%%%%%%%%%%%%%%%%%%%%%%%%%%%%%%%%%
	\section{Introduction}

Extremal black holes play a privileged role at the intersection of classical gravity, quantum field theory, and holography, making them a good area to explore new physics \cite{Sen:2007qy,Compere:2012jk,Horowitz:2023xyl}. Their Hawking temperature vanishes, while they generally have a nonzero entropy. Under mild assumptions on the matter sector, the near‑horizon region decouples from the bulk  and controls the spacetime dynamics \cite{Sen:2007qy,Astefanesei:2006dd,Astefanesei:2007bf,Kunduri:2007vf,Kunduri:2013gce}. In other words, the attractor mechanism fixes the horizon data in terms of conserved charges, largely independent of asymptotic data. In four dimensions, a smooth, stationary, axisymmetric and asymptotically flat extremal solution admits a near‑horizon extremal geometry (NHEG) with an  $SO(2,1)\times U(1)$  isometry, whose dynamics are determined by an entropy functional \cite{Sen:2007qy,Sen:2005iz,Astefanesei:2006dd,Astefanesei:2007bf,Kunduri:2007vf,Kunduri:2013gce}. At the same time, recent analyses caution that smooth extremal horizons are not guaranteed in general \cite{Herdeiro:2025blx,Blazquez-Salcedo:2025cpu,Horowitz:2022mly,Horowitz:2024kcx}. Hence, the would‑be near‑horizon geometry may fail to arise. These considerations motivate a concrete study/construction of attractors in the presence of various matter fields, which is the main goal of the present work.

Scalar fields coupled to electromagnetism arise naturally in different domains of physics \cite{GIBBONS1988741,Garfinkle:1990qj,Sen:1992ua}.  In dimensional reduction, a dilaton appears already in the simplest Kaluza-Klein setup: starting from five-dimensional vacuum Einstein gravity and compactifying on a circle yields four-dimensional Einstein-Maxwell theory coupled to a massless scalar (the dilaton) \cite{Kaluza:1921tu,Klein1926}. String effective actions further supply a pseudoscalar axion via dualization of the Kalb-Ramond 2‑form \cite{PhysRevLett.40.223,PhysRevLett.40.279,Svrcek:2006yi,Ortin:2015hya}. The minimal four‑dimensional truncation that captures these features is the Einstein-Maxwell-dilaton-axion (EMDA) model. From the particle‑physics side, axions arise as pseudo‑Nambu-Goldstone bosons in the context of the strong‑CP problem and are compelling dark‑matter candidates \cite{Marsh:2015xka,Peccei:1977hh,Bergstrom:2009ib}.

In this work, we analyse attractors and extremal rotating black holes in Einstein-Maxwell-axion theory. We show that rotating attractors with axionic hair exist only in the purely electric or purely magnetic charge sectors; regular dyonic attractors are excluded. This emerges both perturbatively and non‑perturbatively from the axion equation in the NHEG. We, therefore, focus on the purely electric sector and construct rotating EMA attractors across representative values of the axion-photon coupling $g_{_{\psi\gamma\gamma}}$. Second, we show that a large subset (but not all) of these rotating attractors extend to asymptotically flat, rotating extremal solutions. In particular, we construct global solutions that interpolate between the electric NHEG and asymptotically flat infinity. Within this branch, the horizon data are fixed by extremizing the EMA entropy function, decoupled from asymptotic moduli, in line with the attractor mechanism \cite{Sen:2007qy,Astefanesei:2006dd,Astefanesei:2007bf}. We also identify a region in parameter space near the static limit in which the extremal, asymptotically flat solutions do not admit a smooth NHEG.

Hence, we emphasize that while NHEG analyses are an efficient way to organize the extremal sector and make horizon‑level statements, the existence of an NHEG does not imply the existence (or uniqueness) of a corresponding global black‑hole solution. Conversely, the absence of a smooth NHEG is a diagnostic of non‑smooth extremal limits. Our construction explicitly exhibits both behaviours: extended branches where horizon data integrate to full spacetimes, and non‑extended branches where extremal solutions lack a smooth near‑horizon limit.

This paper is organized as follows. Section \ref{sec2} defines the EMA model and sets conventions. Section \ref{attractors} develops the near‑horizon (entropy‑function) formulation and proves the purely electric/purely magnetic branching of regular rotating attractors. Section \ref{sec4} turns to the global problem and formulates the boundary conditions, conserved quantities and Smarr relations in EMA. Section \ref{sec7} presents the numerical extremal families, the comparison with NHEGs, and the emergence of the critical point $P$ where smoothness fails. Appendices collect numerical details and checks (including the recovery of the Kerr-Sen attractor in the appropriate EMDA limit \cite{Sen:2007qy,Sen:2005iz,Astefanesei:2006dd}).

\bigskip

	%%%%%%%%%%%%%%%%%%%%%%%%%%%%%%%%%%%%%%%%%%%%%%%%%%%%%%%%%%%%%%%%%%%%%%%%%%%%%%
	\section{The model}
	\label{sec2}

 The Einstein-Maxwell-axion (EMA) model is characterized by the action 
\begin{equation}
\label{action}
S= \frac{1}{4 \pi}\int \left\{\dfrac{1}{4}R\, \boldsymbol{\epsilon}
-\dfrac{1}{8}\, d\psi\wedge\star d \psi
-\, \dfrac{1}{2}\, 
\mathcal{F}\wedge\star\mathcal{F}
-\, g_{_{\psi\gamma\gamma}}\dfrac{\psi}{2}\, 
\mathcal{F}\wedge\mathcal{F}\right\} \ ,
\end{equation}
where $R$ is the Ricci scalar, $\boldsymbol{\epsilon}$ is the spacetime volume, $\mathcal{F}=d\mathcal{A}$ is the Maxwell field strength 2-form, $\mathcal{A}$ is the 1-form gauge potential, $\psi$ is the axion and $g_{_{\psi\gamma\gamma}}$ is the axion-photon coupling \footnote{In our conventions, we have $\mathcal{F}=\dfrac{1}{2}\mathcal{F}_{\mu\nu}\,dx^\mu\wedge dx^\nu, \boldsymbol{\epsilon}_{r\theta\varphi t}=\sqrt{-g}$. }. Varying the action (\ref{action}) with respect to the fields $g_{\mu\nu}$,  $\mathcal{A}$  {and  $\psi$}  gives the corresponding equations of motion,
\begin{eqnarray}
\label{EME}
E_{\mu\nu}=R_{\mu\nu}-\dfrac{g_{\mu\nu}}{2}R-2T_{\mu\nu}&=&0\,,
\\
\label{ME}
d\left(\star\mathcal{F}+g_{_{\psi\gamma\gamma}}\ \psi\ \mathcal{F}\right)&=&0
\,,
\\
\label{DE}
d\left(\star d\psi\right)-2g_{_{\psi\gamma\gamma}}\ \mathcal{F}\wedge\mathcal{F}&=&0 \,,
 \end{eqnarray}
  {where \(R_{\mu\nu}\) is the Ricci tensor and \(T_{\mu\nu}\) is}  the energy-momentum tensor,
\begin{equation}
\label{tik}
T_{\mu\nu}=
\left(
\mathcal{F}_\mu\,^\sigma\mathcal{F}_{\nu\sigma}-\frac{1}{4}g_{\mu\nu}\mathcal{F}_\sigma\,^\tau \mathcal{F}^\sigma\,_\tau
\right)
+ \dfrac{1}{4}\nabla_\mu\psi\nabla_\nu\psi-\dfrac{1}{8}g_{\mu\nu}\nabla_\tau\psi\nabla^{\tau}\psi\,.
\end{equation}

In this convention, when the dilaton field is included and $g_{_{\psi\gamma\gamma}}=1$, one obtains the low-energy effective
theory describing the heterotic string, while for $g_{_{\psi\gamma\gamma}}=0$ it reduces to the Einstein-Maxwell theory. Static, spherically symmetric black holes in the model \eqref{action}, in the purely electric or magnetic sector, have a trivial axion field (assuming a constant value) for any coupling $g_{_{\psi\gamma\gamma}}$. On the other hand, dyonic solutions allow for a nontrivial axion field. Solutions have been studied both perturbatively and numerically \cite{Lee:1991jw,Balakin:2017nbg,Nakarachinda:2025bvy,Fernandes:2019kmh}. Rotation allows for a nontrivial axion field even in the purely electric sector, for instance. Perturbative solutions have been studied in \cite{Boskovic:2018lkj,KIczek:2021vlc}, while non-extremal black holes were numerically studied in \cite{Burrage:2023zvk}.

We are interested in asymptotically flat, stationary and axisymmetric solutions. Such spacetimes admit two Killing vector fields, which can be written in adapted coordinates as
\(\xi = \partial_t\) and \(\eta = \partial_\varphi\), where \(t\) and \(\varphi\) denote, respectively, the asymptotic time and the azimuthal angle. Since we consider asymptotically flat configurations, the two Killing fields commute, \([\xi,\eta]=0\), without loss of generality \cite{Carter:1970ea}. An important simplification is achieved by noticing that the circularity condition is an imposition of the field equations. The proof is outside the scope of this paper, but it can be achieved following standard approaches \cite{heusler_1996,Kundt:1966zz,Carter:1969zz,Carter:2009nex,Bokulic:2023oxw,Herdeiro:2025blx,Straumann:2013spu}. Hence, the line element can be expressed as
\begin{equation}
\label{m1}
d s^2=-\frac{\rho^2}{X(\rho, z)} d t^2+X(\rho, z)[d \varphi-w(\rho, z) d t]^2+\frac{e^{2 h(\rho, z)}}{X(\rho, z)}\left[d \rho^2+d z^2\right]\,.
\end{equation}

Given a Killing vector \(\kappa\), we define its twist 1-form
\(\boldsymbol{\omega} = \star(\kappa \wedge d\kappa)\).  
The twist associated with \(\kappa\) obeys
\begin{equation}\label{domega}
 d\boldsymbol{\omega}=-2\iota_\kappa\star R(\kappa)=4 \iota_\kappa\mathcal{F}\land\iota_\kappa\left(\star\mathcal{F}\right)\,.
\end{equation}

We introduce electric and magnetic potentials associated with a Killing vector \(\kappa\) through
\begin{equation}
\label{electr_potentials}
    d\Phi=-\iota_\kappa \mathcal{F}\,,\qquad d\Psi=\iota_\kappa\left(\star\mathcal{F}+g_{_{\psi\gamma\gamma}}\,\psi \mathcal{F}\right)\,.
\end{equation}

Using these definitions, Eq.~\eqref{domega} can be rewritten as
\begin{equation}
    d\left(\boldsymbol{\omega}+2\Phi d\Psi-2\Psi d\Phi\right)=0\,.
\end{equation}

Therefore, there exists a scalar potential \(\chi\) such that $d\chi=\boldsymbol{\omega}+2\Phi d\Psi-2\Psi d\Phi$. 
Following the approach of \cite{Galtsov:1995zm,Yazadjiev:2010bj,Wells:1998gc}, we now perform a dimensional reduction by choosing \(\kappa = \eta\), the spacelike axial Killing vector.\footnote{Here the reduction is carried out with respect to the axial Killing vector in order to obtain a positive-definite metric on the target space \cite{Yazadjiev:2010bj}. In contrast, in \cite{Galtsov:1995zm} the reduction is performed along the timelike Killing vector.}  With this choice, the field equations reduce to the following system on the auxiliary two-dimensional flat space with coordinates \((\rho,z)\), where
\(\bar{\nabla}=(\pa_\rho\;,\;\pa_z)\) and \(\bar{\nabla} U \cdot \bar{\nabla} V = \partial_\rho U\,\partial_\rho V + \partial_z U\,\partial_z V\)
\begin{subequations}\label{eq:dimred}
\begin{align}
\label{eq:dimred-a}
& \rho^{-1}\bar{\nabla}\cdot\left(\rho \bar{\nabla}X\right)
 = X^{-1}\left(\bar{\nabla}X\right)^2
 - X^{-1}\left(\bar{\nabla}\chi - 2\Phi \bar{\nabla}\Psi
               + 2\Psi \bar{\nabla}\Phi\right)^2
\\\nonumber
& \quad
 - 2\left(\bar{\nabla}\Phi\right)^2
 - 2\left(\bar{\nabla}\Psi
          + g_{_{\psi\gamma\gamma}}\,\psi\,\bar{\nabla}\Phi\right)^2,
\\[10pt]
\label{eq:dimred-b}
& \bar{\nabla}\cdot\left[\rho X^{-2}
   \left(\bar{\nabla}\chi - 2\Phi \bar{\nabla}\Psi
         + 2\Psi \bar{\nabla}\Phi\right)\right] = 0,
\\[10pt]
\label{eq:dimred-c}
& \rho^{-1}\bar{\nabla}\cdot\left(\rho X^{-1}\bar{\nabla}\Phi\right)
 = -\left(\bar{\nabla}\Psi
          + g_{_{\psi\gamma\gamma}}\,\psi\,\bar{\nabla}\Phi\right)\cdot
   \left[X^{-2}\left(\bar{\nabla}\chi - 2\Phi \bar{\nabla}\Psi
                     + 2\Psi \bar{\nabla}\Phi\right)
        + X^{-1}\bar{\nabla}\psi\right],
\\[10pt]
\label{eq:dimred-d}
& \rho^{-1}\bar{\nabla}\cdot\left(\rho X^{-1}\bar{\nabla}\Psi\right)
 = -\bar{\nabla}\Phi\cdot
   \left[g_{_{\psi\gamma\gamma}} X^{-1}\bar{\nabla}\psi
        - X^{-2}\left(\bar{\nabla}\chi - 2\Phi \bar{\nabla}\Psi
                     + 2\Psi \bar{\nabla}\Phi\right)\right]
\\\nonumber
& \quad
 +\left(\bar{\nabla}\Psi
        + g_{_{\psi\gamma\gamma}}\,\psi\,\bar{\nabla}\Phi\right)\cdot
   \left[g_{_{\psi\gamma\gamma}}\,\psi\,X^{-2}
         \left(\bar{\nabla}\chi - 2\Phi \bar{\nabla}\Psi
              + 2\Psi \bar{\nabla}\Phi\right)
        + g_{_{\psi\gamma\gamma}}^{2}\,\psi\,X^{-1}\bar{\nabla}\psi\right],
\\[10pt]
\label{eq:dimred-e}
& \rho^{-1}\bar{\nabla}\cdot\left(\rho \bar{\nabla}\psi\right)
 = 4 g_{_{\psi\gamma\gamma}} X^{-1}
   \bar{\nabla}\Phi\cdot
   \left(\bar{\nabla}\Psi
        + g_{_{\psi\gamma\gamma}}\,\psi\,\bar{\nabla}\Phi\right).
\end{align}
\end{subequations}

These equations can be viewed as the field equations on a two-dimensional manifold for a set of five real scalar fields
\begin{equation}
\varphi^A = (X,\chi,\Phi,\Psi,\psi)\,, \qquad A = 1,\dots,5\,,
\end{equation}
which define  a target manifold \(\mathcal{N}\) equipped with a Riemannian metric \(G\)
\begin{equation}
dL^2 = G_{AB}\,dX^A dX^B
= \frac{dX^2 + \bigl(d\chi + 2\Psi\,d\Phi - 2\Phi\,d\Psi\bigr)^2}{X^2}
  + \frac{4}{X}\left[d\Phi^2 
    + \bigl(d\Psi + g_{_{\psi\gamma\gamma}}\,\psi\,d\Phi\bigr)^2\right]
  + d\psi^2 \,.
\end{equation}

Hence, the stationary, axisymmetric sector of the EMA model can be obtained equivalently from the sigma-model action
\begin{equation}
\label{eq:sigma-action}
S_\sigma
 = \int \Bigl[ G_{AB}(\varphi)\,\nabla_i\varphi^A\,\nabla_j\varphi^B\,h^{ij}\Bigr]\sqrt{h}\,d^2x\,,
\end{equation}
where \(h_{ij}\) is the metric on the two-dimensional orbit space. It is natural to ask whether this sigma model admits hidden symmetries, as in the Einstein-Maxwell and certain Einstein-Maxwell-dilaton(-axion) cases where the target space is symmetric, \(\nabla_E R_{ABCD} = 0\), and these symmetries can be used to construct solution-generating techniques. This occurs, for instance, in EMDA with the stringy coupling and in EMD with the Kaluza-Klein coupling \cite{Galtsov:1995zm,Yazadjiev:2010bj,Wells:1998gc}. For the EMA model considered here, however, explicit computation shows that this condition fails for generic nonzero axion-photon coupling \(g_{_{\psi\gamma\gamma}} \neq 0\), so the target space is not symmetric.

\bigskip

In the next section, we take the near‑horizon (rotating attractor) ansatz appropriate to extremal, axisymmetric solutions and analyze the coupled ordinary differential equations (ODEs) following the entropy‑function   {formalism}. As we will see, the axion equation forces regular solutions to be purely electrically or purely magnetically charged, a point we establish perturbatively and non‑perturbatively.

%%%%%%%%%%%%%%%%%%%%%%%%%%%%%%%%%%%%%%%%%%%%%%%%%%%%%%%%%%%%%%%%%%%%%%%%%%%%%%%%%%%%%%%%%%%%%%%%%%%
\section{Attractors}\label{attractors}

Any smooth, stationary, axisymmetric, asymptotically flat extremal 4D black hole, such as those described by \eqref{action}, admits a NHEG whose isometry group is $SO(2,1)\times U(1)$ \cite{Reall:2002bh,Kunduri:2007vf}. The generic ansatz for the metric and matter fields can be taken in the form \cite{Kunduri:2007vf,Astefanesei:2006dd,Astefanesei:2007bf}
\begin{eqnarray} 
\label{metric}
ds^2= v_1(\theta) 
\left(
      -R^2 dT^2+\frac{dR^2}{R^2}+\beta^2 d\theta^2
\right)
+v_2(\theta)  (d \tilde \varphi+K R dT)^2
\end{eqnarray} 
\begin{eqnarray} 
\label{matter-nh}
\mathcal{A}=b (\theta) (d \tilde \varphi+ K R dT)+q R dT,~~\psi \equiv \psi(\theta)~,
\end{eqnarray}
where $q$, $K$ and $\beta$ are real constants, and $v_{1}$, $v_{2}$, $b$ and $\psi$ are functions of $\theta$.

Although we ultimately wish to associate the NHEG   {with}  a parent extremal black hole, such a spacetime exists   {in its own right;}  in particular, it   {is an exact solution of the field equations}  for all finite   {\(R\)}, not just for small   {\(R\) \cite{Sen:2007qy}}. Considering the NHEG as a geometry by itself, the equations of motion can be obtained by extremizing the corresponding action. Moreover, one can introduce a new functional \cite{Sen:2005iz,Astefanesei:2006dd}, called the entropy function and defined by\footnote{Here, we follow the approach as first proposed in \cite{Sen:2005iz,Astefanesei:2006dd}, but an equivalent derivation can also be carried out using  {an approach closely related to}  the Iyer-Wald entropy construction \cite{Compere:2015mza,Compere:2015bca,Hajian:2013lna}.}
\begin{equation}
    \mathcal{E}\!\bigl[J,Q_e,K,\beta,q,
                 v_{1}(\theta),v_{2}(\theta),
                \psi(\theta),b(\theta)\bigr]
  \;=\;
  2\pi\!\left(
          J K
          + Q_e\, q -\int d \tilde{\varphi}\,d \theta\,\sqrt{-g}\,\mathcal{L}
        \right)\,, \label{entropy_function}
 \end{equation}
where $Q_e$ and $J$ are related to the electric charge and angular momentum, respectively. As shown in   {\cite{Astefanesei:2006dd,Sen:2005iz,Astefanesei:2007bf}}, the entropy and the near-horizon background of a rotating extremal black hole are obtained by extremizing the entropy function of the near-horizon parameters and charges.   {Then, extremizing the entropy function \eqref{entropy_function},}  the equations of motion take the form
\begin{equation}
    \frac{\partial\mathcal{E}}{\partial K}=0,\quad
\frac{\partial\mathcal{E}}{\partial\beta}=0,\quad
\frac{\partial\mathcal{E}}{\partial q}=0,\quad
\frac{\delta\mathcal{E}}{\delta v_{1}(\theta)}=0,\quad
\frac{\delta\mathcal{E}}{\delta v_{2}(\theta)}=0,\quad
\frac{\delta\mathcal{E}}{\delta \psi(\theta)}=0,\quad
\frac{\delta\mathcal{E}}{\delta b(\theta)}=0.
\label{varE}
\end{equation}

Moreover, the entropy of the parent black hole coincides with the value of the entropy function in its extremum   {\cite{Astefanesei:2006dd,Sen:2005iz,Astefanesei:2007bf}}
\begin{equation}
    S_{\text{BH}}=\mathcal{E}\,.
\end{equation}

While $J$ and $Q_e$ can be obtained by the extremization of $\mathcal{E}$ with respect to $K$ and $q$, respectively, the magnetic charge can be directly read from the boundary of $b$\footnote{We emphasize that the angular momentum, electric and magnetic charge can be equivalently obtained using definitions \eqref{ang_komar} and \eqref{Qe_QM} with metric \eqref{metric}.}
\begin{equation} \label{magnetic_charge}
     Q_m=\dfrac{b(\pi)-b(0)}{2}\,.
\end{equation}

Let us remark that although the entropy function offers another formalism to compute the entropy of an extremal black hole if the latter solution exists, it neither guarantees the existence nor the uniqueness  of   such an extremal black hole  solution. This means that the analysis does not tell whether the full black hole solution, interpolating between $AdS^2 \times S^1$ near-horizon geometry and the asymptotically flat Minkowski space, really exists.  Moreover, since uniqueness need not hold, a single NHEG may   arise as the near-horizon limit of multiple different extremal black hole spacetimes, \emph{i.e.} their near-horizon geometries may coincide even though their bulk configurations differ away from the horizon. Conversely, if the extremal black hole  spacetime  is smooth close to the horizon, the NHEG should exist  {\cite{Kunduri:2007vf,Kunduri:2013gce}}. Hence, we first aim to construct the NHEG and then investigate which of these near-horizon solutions extend to regular, asymptotically flat extremal black holes.

We obtain the equations of motion by extremizing the entropy function   {\eqref{entropy_function}--\eqref{varE}; equivalently, the same system follows directly from the Einstein and matter equations evaluated on the ansatz \eqn{metric}--\eqn{matter-nh} (see Appendix~\ref{sec:numerics-attractor} and}  \cite{Astefanesei:2006dd}).   {It follows that}  the metric functions $v_1$ and $v_2$ are not independent, but satisfy
 \begin{equation}
v_2(\theta)=\frac{s^2\sin^2 \theta}{v_1(\theta)}\,,
\end{equation}
where the physical   {interpretation}  of the constant $s$ is   {discussed}  in Appendix~\ref{sec:numerics-attractor}. It is also convenient to introduce the  coordinate $u=\cos\theta\in[-1,1]$, {in terms of which}  the equations of motion   {become}
\begin{subequations}\label{eq:NH-EMA}
\allowdisplaybreaks
\begin{align}
v_{1}'' &=
 \frac{v_{1}^{2} b'\, ^{2}}{s^{2}\mathcal{D}}
 +\frac{\mathcal{B}^{2}+u\,v_{1}'}{\mathcal{D}}
 +\frac{3\,(K^{2}s^{2}+v_{1}'\,^{2})}{4v_{1}}
 -\frac{v_{1}}{\mathcal{D}}
 -\frac{1}{4}v_{1}\psi'\,^{2}\label{eq:NH-EMA-v1},\\[4pt]
b'' &= -\frac{v_{1} b' v_{1}' - g_{_{\psi\gamma\gamma}}\,s\,v_{1}\psi'\,\mathcal{B}
               +K^{2}s^{2}\, b + K q s^{2}}{v_{1}^{2}},\\[4pt]
\psi'' &= \frac{-\,4 g_{_{\psi\gamma\gamma}}\, b'\,\mathcal{B} + 2 s u\,\psi'}{s\,\mathcal{D}},\label{eq:NH-EMA-phi}\\[4pt]
0 &= -\frac{4 b'\,^{2}}{s^{2}}
     +\frac{-4\mathcal{B}^{2}-4u\,v_{1}'}{v_{1}^{2}}
     -\frac{\mathcal{D}\,(K^{2}s^{2}+v_{1}'\,^{2})}{v_{1}^{3}}
     +\frac{4 - \mathcal{D}\,\psi'\,^{2}}{v_{1}}\,,\label{eq:NH-EMA-cons}
\end{align}
\end{subequations}
where we have defined $\mathcal{D}(u)\coloneqq 1-u^{2}, \mathcal{B}(u)\coloneqq K\,b(u)+q$ and prime denotes derivative with respect to $u$. In the case where $g_{_{\psi\gamma\gamma}}=0$, the above system admits the near-horizon extremal Kerr-Newman (NHEKN) \cite{Newman:1965my,Bardeen:1999px}  as solution 
\begin{equation}
  \begin{aligned}\label{NHEKNDEF}
v_{1}^{\scriptscriptstyle\mathrm{{KN}}}(u) &= a^{2}(1+u^{2}) + Q_{e}^{2} + Q_{m}^{2},\qquad
b^{\scriptscriptstyle\mathrm{KN}}(u) = \frac{a Q_{e}(1-u^{2})\,M \; -\; 2 a^{2} Q_{m} u \; -\; Q_{m} u (Q_{e}^{2}+Q_{m}^{2})}{a^{2}(1+u^{2}) + Q_{e}^{2} + Q_{m}^{2}},\\
\psi^{\scriptscriptstyle\mathrm{KN}}(u) &= 0,\qquad
M = \sqrt{a^{2}+Q_{e}^{2}+Q_{m}^{2}},\\
s^{\scriptscriptstyle\mathrm{KN}}&= 2 a^{2}+Q_{e}^{2}+Q_{m}^{2},\qquad
K^{\scriptscriptstyle\mathrm{KN}}= \frac{2 a M}{2 a^{2}+Q_{e}^{2}+Q_{m}^{2}},\qquad
q^{\scriptscriptstyle\mathrm{KN}}= \frac{Q_{e}(Q_{e}^{2}+Q_{m}^{2})}{2 a^{2}+Q_{e}^{2}+Q_{m}^{2}}\,,
\end{aligned} 
\end{equation}
with $a,Q_e,Q_m>0$ arbitrary fixed parameters.

%%%%%%%%%%%%%%%%%%%%%%%%%%%%%%%%%%%%%%%%%%%%%%%%%%%%%%%%%%%%%%%%%%%%%%%%%%%%%%%%%%%%%%
\subsection{Perturbative analysis}\label{sec:perturbar}

We organize the perturbative construction around two analytic backgrounds. As mentioned previously, when $g_{_{\psi\gamma\gamma}}=0$, the NHEKN is a solution of the system \eqref{eq:NH-EMA}. Equivalently, near-horizon extremal Kerr (NHEK) is a solution of \eqref{eq:NH-EMA} in the absence of electric/magnetic charge. Hence, one can attempt to construct a perturbative solution around these solutions.
\begin{eqnarray} 
&&
v_1(u)= \sum_{ k\geq 0} \epsilon^k v_{1k}(u),~~
b (u)= \sum_{ k\geq 0} \epsilon^k 
b_{k}(u),~~
\psi (u)= \sum_{ k\geq 0} \epsilon^k \psi_k(u),~~
 \\
 \nonumber
 &&
{\rm and}~~~
K= \sum_{ k\geq 0} \epsilon^k  K_k ,~~
 q= \sum_{ k\geq 0} \epsilon^k q_k ,~~
\end{eqnarray}  
with $\epsilon$ an infinitesimal parameter, the equations can easily be solved order by order. In Secs. \ref{NHEKN} and \ref{NHEK}, the zeroth order is chosen to be the NHEKN and NHEK, respectively. Such that, to order $\mathcal{O}(\epsilon^0)$, the equations \eqref{eq:NH-EMA} are exactly satisfied.

Notice that, a priori, one would have to consider a perturbative expansion for the parameter \(s\) as well. However, the action \eqref{action} is invariant under a global rescaling of the fields. Consequently, the NHEG is invariant under
\begin{equation}
        v_{1}\rightarrow \lambda^{2}v_{1},\qquad
        s\rightarrow \lambda^{2}s,\qquad
        b\rightarrow \lambda b,\qquad
        q\rightarrow \lambda q ,
\end{equation}
with \(K\), \(u\) and \(\psi\) left unchanged. Therefore, solutions related by this transformation describe the same dimensionless attractor data, differing only by the overall NHEG scale. We can thus fix \(s\) throughout the perturbative construction, without loss of generality. This fixes the boundary condition for $v_{1k}$ with $k\ge 1$ to be, at each order\footnote{We also checked that performing the computation with perturbations of \(s\) included, and imposing only the physical pole condition \(s_k=v_{1k}(\pm1)\) as a consequence of \eqref{new-cond}, leads to the same final results.}
\begin{equation}
    v_{1k}(-1)=v_{1k}(1)=0\,.
\end{equation}

Notice also that the system is invariant under the residual gauge symmetry
\begin{equation}\label{eq:gauge}
    b\rightarrow b+\Lambda,\qquad q\rightarrow q-K\Lambda ,
\end{equation}
with \(\Lambda\) a constant. We fix this gauge freedom by imposing, at each order,
\begin{equation}
    b_k(-1)+b_k(1)=0\,.
\end{equation}

 As we shall see in the following perturbative analysis, a number of integration constants in the expression of
 the higher order terms
are fixed by imposing that the solution is smooth everywhere (in particular without conical singularities).
Also, for the scalar field, we impose
$\psi=0$
if the electromagnetic field trivializes.
Then, 
at  each order $k>0$ in perturbation theory,
there is a single free integration constant, $B_k$,
which fixes the $k$-th order contribution to the total
electric/magnetic charge.

 %%%%%%%%%%%%%%%%%%%%%%%%%%%%%%%%%%%%%%%%%%%%%%%%%%%%%%%
 \subsubsection{Perturbation around NHEKN}\label{NHEKN}

 Treating the axion-photon coupling as a small parameter $\epsilon=g_{_{\psi\gamma\gamma}}$, we construct a perturbative expansion around the NHEKN background ($g_{_{\psi\gamma\gamma}}=0$). 
\begin{align}
 v_1 &= v_1^{\scriptscriptstyle\mathrm{KN}}+g_{_{\psi\gamma\gamma}} v_{11}+\mathcal{O}(g_{_{\psi\gamma\gamma}}^2),&
 b &= b^{\scriptscriptstyle\mathrm{KN}}+g_{_{\psi\gamma\gamma}} b_{1}+\mathcal{O}(g_{_{\psi\gamma\gamma}}^2),\nonumber\\
 K &= K^{\scriptscriptstyle\mathrm{KN}}+g_{_{\psi\gamma\gamma}} K_{1}+\mathcal{O}(g_{_{\psi\gamma\gamma}}^2),&
 q &= q^{\scriptscriptstyle\mathrm{KN}}+g_{_{\psi\gamma\gamma}} q_{1}+\mathcal{O}(g_{_{\psi\gamma\gamma}}^2),\nonumber\\
 \psi &= g_{_{\psi\gamma\gamma}}\psi_{1}+\mathcal{O}(g_{_{\psi\gamma\gamma}}^2)
\end{align}

Although we have not been able to   {extend the perturbative solution}  beyond first order  in $g_{_{\psi\gamma\gamma}}$ in closed form, the leading  axion correction already exhibits a nontrivial feature.
\begin{equation}
\begin{aligned}
\psi_1\;=\;&C_{2}
+\frac{1}{\bigl(2a^{2}+Q_m^{2}+Q_e^{2}\bigr)^{2}}
\Bigg\{
-\,\frac{2\bigl(2a^{2}+Q_m^{2}+Q_e^{2}\bigr)}{Q_m^{2}+Q_e^{2}+a^{2}(1+u^{2})}
\Bigl[
2a^{2}Q_m Q_e + 2Q_m Q_e\bigl(Q_m^{2}+Q_e^{2}\bigr)
\\
&\qquad\qquad\qquad\qquad
+\,a\bigl(-Q_m^{2}+Q_e^{2}\bigr)\sqrt{a^{2}+Q_m^{2}+Q_e^{2}}\;u
\Bigr]
+2\bigl(Q_m^{4}-Q_e^{4}\bigr)
\arctan\!\Bigl(\frac{a u}{\sqrt{a^{2}+Q_m^{2}+Q_e^{2}}}\Bigr)
\\
&\qquad
-\Bigl[
4a\bigl(Q_m^{2}-Q_e^{2}\bigr)\sqrt{a^{2}+Q_m^{2}+Q_e^{2}}
+\bigl(2a^{2}+Q_m^{2}+Q_e^{2}\bigr)^{2}C_{1}
\Bigr]\operatorname{arctanh}u
\\
&\qquad
+\,2Q_m Q_e\bigl(Q_m^{2}+Q_e^{2}\bigr)
\Bigl[
\log\!\bigl(Q_m^{2}+Q_e^{2}+a^{2}(1+u^{2})\bigr)
-\log(1-u^{2})
\Bigr]
\Bigg\}.
\end{aligned}
\end{equation}

Here $C_1$ and $C_2$ are integration constants, and $(a,Q_e,Q_m)$ are the NHEKN rotation and charge parameters. Although the explicit form of $\psi_1$ is cumbersome, a quick inspection shows that it has a potential  singular structure. The $\operatorname{arctanh}u$ and $\log(1-u^{2})$ terms encode potential logarithmic divergences at the poles $u=\pm 1$.   By expanding the solution in series around $u=\pm 1$, one sees that the free constant  $C_1$   {can be chosen}  to cancel the divergence at either $u=1$ or $u=-1$, but not at both poles simultaneously for generic dyonic data   {$Q_eQ_m\neq 0$}. More explicitly, defining
\begin{align}
 X &= 4a(Q_m^2-Q_e^2)\sqrt{a^2+Q_m^2+Q_e^2}+(s^{\scriptscriptstyle\mathrm{KN}})^2 C_1,\nonumber\\
 Y &= 2Q_mQ_e(Q_m^2+Q_e^2),
\end{align}
one obtains the singular parts
\begin{align}
 \psi_{1}
 &=\frac{1}{(s^{\scriptscriptstyle\mathrm{KN}})^2}\left(\frac{X}{2}- Y\right)\log(1-u)+\mathcal{O}(1),
 \qquad u\rightarrow 1,\nonumber\\
 \psi_{1}
 &=\frac{1}{(s^{\scriptscriptstyle\mathrm{KN}})^2}\left(-\frac{X}{2}-Y\right)\log(1+u)+\mathcal{O}(1),
 \qquad u\rightarrow -1 .
\end{align}
Regularity at $u=1$ requires $X=2Y$, while regularity at $u=-1$ requires $X=-2Y$. These two conditions are compatible if and only if $X=Y=0$. The condition $Y=0$ implies
\begin{equation}
 Q_mQ_e(Q_m^2+Q_e^2)=0.
\end{equation}

Consequently, the first order solution can be made everywhere regular when the background is purely electric or purely magnetic. In the purely electric case, $Q_m=0$, the remaining condition $X=0$ fixes the integration constant as
\begin{equation}
    C_1=\frac{4 a Q_e^2 \sqrt{a^2+Q_e^2}}{\left(2 a^2+Q_e^2\right)^2}\,,
\end{equation}
leading the first order axion solution to be of the form
\begin{equation}
 \psi_1=  C_2 -\frac{2 a Q_e^2 u \sqrt{a^2+Q_e^2}}{\left(2 a^2+Q_e^2\right) \left(a^2 \left(u^2+1\right)+Q_e^2\right)}-\frac{2 Q_e^4 \tan ^{-1}\left(\frac{a u}{\sqrt{a^2+Q_e^2}}\right)}{\left(2 a^2+Q_e^2\right)^2}\,.
\end{equation}

The remaining constant \(C_2\) is then fixed by the chosen additive convention for the
axion, for instance, setting the axion to zero if $Q_e=0$.

This perturbative analysis around NHEKN therefore already signals that a smooth axionic attractor is incompatible with dyonic configurations. In what follows, we will see that dyonic solutions are also ruled out in the background of Kerr.
 %%%%%%%%%%%%%%%%%%%%%%%%%%%%%%%%%%%%%%%%%%%%%%%%%%%%%%%
 \subsubsection{Perturbation around NHEK}\label{NHEK}

We now construct perturbative solutions around the NHEK background.  Imposing $Q_e=Q_m=0$  {in Eq. \eqref{NHEKNDEF}}  yields the NHEK background
\begin{equation}
   v_{10}(u) = a^2(1 + u^2), \quad b_0(u) = 0, \quad \psi_0(u) = 0, \quad K_0 = 1\,.
\end{equation}

Following the same perturbative expansion in powers of $\epsilon$, but now with $\epsilon$ an independent small parameter and $g_{_{\psi\gamma\gamma}}$ held fixed at a generic value, we have solved the ODE system \eqref{eq:NH-EMA} up to eighth order. For brevity, we only display the results up to fourth order. It is worth detailing how the solution at each order is constructed and how the integration constants are fixed, since this clarifies the iterative structure of the expansion and the role played by the remaining integration constants. 

At order $k\geq 1$, the expansion in powers of $\epsilon$ turns \eqref{eq:NH-EMA} into an ODE system for $(v_{1k},b_k,\psi_k)$ whose source terms depend on the lower-order solutions $(v_{1j},b_j,\psi_j,K_j,q_j)$ with $j<k$; the orders are therefore solved recursively. At a given order, the three second order equations for $(v_{1k},b_k,\psi_k)$ give six integration constants. Together with the two parameters $K_k$ and $q_k$, this amounts to eight constants. These are constrained by the scale-fixing condition $v_{1k}(\pm1)=0$, the gauge-fixing condition $b_k(-1)+b_k(1)=0$, the convention that the axion vanishes when the electromagnetic field trivializes, the constraint equation in \eqref{eq:NH-EMA}, and regularity of $\psi_k$ at the poles $u=\pm1$. These six conditions leave two constants, which would in principle allow for a dyonic perturbative solution. 

As in the NHEKN analysis of Sec.~\ref{NHEKN}, however, regularity at a single order does not guarantee regularity of the full tower. If the lower order perturbative solutions are dyonic, the subsequent order inevitably develops logarithmic divergences at $u=\pm1$ that cannot be removed at both poles simultaneously by any choice of the remaining constants. Imposing regularity order by order thus propagates back to the lower orders and forces them to be purely electric or purely magnetic, setting one of the two constants to zero and leaving a single integration constant $B_k$, which fixes the $k$-th order contribution to the (would-be) electric (or magnetic) charge. In what follows we present the purely electric branch.

The axion solution reads
\begin{equation}
\begin{aligned}
\psi_{1} &= 0,\qquad
\psi_{2} = -\frac{B_{1}^{2}\,g_{_{\psi\gamma\gamma}}\,u}{a^{2}(u^{2}+1)},\qquad
\psi_{3} = -\frac{2 B_{1} B_{2}\,g_{_{\psi\gamma\gamma}}\,u}{a^{2}(u^{2}+1)},\\
\psi_{4} &= \frac{g_{_{\psi\gamma\gamma}}}{6 a^{4}}\!\left[
-\frac{6 a^{2} u\,\bigl(2 B_{1} B_{3}+B_{2}^{2}\bigr)}{u^{2}+1}
-\frac{B_{1}^{4}\bigl(g_{_{\psi\gamma\gamma}}^{2}-3\bigr)\,u\,(u^{2}+3)}{(u^{2}+1)^{2}}
+ 3 B_{1}^{4}\bigl(g_{_{\psi\gamma\gamma}}^{2}-1\bigr)\arctan u
\right]\!.
\end{aligned}
\end{equation}

The solution for the gauge field is
\begin{equation}
\begin{aligned}
b_{1} &= B_{1}\,\frac{1-u^{2}}{u^{2}+1},\qquad
b_{2} = B_{2}\,\frac{1-u^{2}}{u^{2}+1},\qquad  b_{3} = \frac{(1-u^{2})\!\left[\,3 a^{2} B_{3}(u^{2}+1) + B_{1}^{3}\bigl(g_{_{\psi\gamma\gamma}}^{2}-3\bigr)\right]}{3 a^{2}(u^{2}+1)^{2}}\,,\\
b_{4} &= \frac{(1-u^{2})\!\left[\,a^{2} B_{4}(u^{2}+1) + B_{1}^{2} B_{2}\bigl(g_{_{\psi\gamma\gamma}}^{2}-3\bigr)\right]}{a^{2}(u^{2}+1)^{2}}.
\end{aligned}
\end{equation}

The metric function
\begin{equation}
\begin{aligned}
v_{11} &= 0,\qquad
v_{12} = \tfrac{1}{2} B_{1}^{2}(1-u^{2}),\qquad
v_{13} = B_{1} B_{2}(1-u^{2}),\\
v_{14} &= \frac{(1-u^{2})\!\left[
24 a^{2} B_{1} B_{3}(u^{2}+1) + 12 a^{2} B_{2}^{2}(u^{2}+1)
+ B_{1}^{4}\!\left(3\bigl(g_{_{\psi\gamma\gamma}}^{2}-4\bigr) u^{2} + 5 g_{_{\psi\gamma\gamma}}^{2}-12\right)
\right]}{24 a^{2}(u^{2}+1)}.
\end{aligned}
\end{equation}

Finally, the solution parameters are
\begin{equation}
K_{1}=K_{2}=K_{3}=0,\qquad
K_{4}=\frac{B_{1}^{4}\bigl(2 g_{_{\psi\gamma\gamma}}^{2}-3\bigr)}{24 a^{4}}.
\end{equation}

\begin{equation}
q_{1}=q_{2}=0,\qquad
q_{3}=\frac{B_{1}^{3}\bigl(3-2 g_{_{\psi\gamma\gamma}}^{2}\bigr)}{6 a^{2}},\qquad
q_{4}=\frac{B_{1}^{2} B_{2}\bigl(3-2 g_{_{\psi\gamma\gamma}}^{2}\bigr)}{2 a^{2}}.
\end{equation}

 {From the perturbative solution, one may also extract the corresponding electric charge and
angular momentum. In the purely electric branch, they take the form}
\begin{align}
Q_e &= B_1 \,\epsilon + B_2 \,\epsilon^2 + \epsilon^3 \left[
B_3 + \frac{B_1^3}{6 a^2}\left(g_{_{\psi\gamma\gamma}}^{\,2}-3\right)
\right]
\nonumber\\
&\quad + \epsilon^4 \left[
B_4 + \frac{B_1^2 B_2}{2 a^2}\left(g_{_{\psi\gamma\gamma}}^{\,2}-3\right)
\right]
+ \mathcal{O}(\epsilon^5) \, ,
\\[0.5em]
J &= a^2 + \frac{B_1^4}{24 a^2}\Bigl(-3 + 2 g_{_{\psi\gamma\gamma}}^{\,2}\Bigr)\epsilon^4
+ \mathcal{O}(\epsilon^5) \, .
\end{align}
In particular, the first correction to the angular momentum appears only at
$\mathcal{O}(\epsilon^4)$, whereas the electric charge is fixed order by order by the
integration constants $B_1$, $B_2$, $B_3$ and $B_4$. Finally, up to fourth order, the entropy
of the solutions can be expressed as
\begin{equation}
    S_{BH}=2\pi J\left[1+\dfrac{3-2 g_{_{\psi\gamma\gamma}}^2}{24}\left(\dfrac{Q_e^2}{J}\right)^2\right]+\mathcal{O}(\epsilon^5)\,.
\end{equation}

 These expressions complete our perturbative construction around NHEK. 
The regular solutions are fully characterized, order by order, by the coefficients $B_k$, 
which fix the electric charge. In the next subsection, we will show that the absence of rotating dyonic attractors is not an artifact of the perturbative expansion, but follows from a simple non-perturbative integral constraint on the axion equation.

%%%%%%%%%%%%%%%%%%%%%%%%%%%%%%%%%%%%%%%%%%%%%%%%%%%%%%%
\subsection{Necessary condition}

Now, we will see that the non existence of dyonic perturbative solutions is not an accident of the perturbation theory, but rather a requirement of the system. Let us consider the attractor equation for the axion
\begin{equation}
0=-\frac{4 g_{_{\psi\gamma\gamma}} \,\big(q + K b(u)\big)\, b'(u)}{s}
+ 2u\,\psi'(u)
- \big(1 - u^2\big)\,\psi''(u).
\end{equation}

Differently from the previous Section \ref{sec:perturbar}, we use the gauge freedom \eqref{eq:gauge} to set the constant $q$ to zero. By regrouping the equation and taking its primitive, we obtain
\begin{equation}
- \frac{2 g_{_{\psi\gamma\gamma}} K}{s} b(u)^2
- (1 - u^2)\,\psi'(u)
= ct\,,
\end{equation}
where $ct$ is a constant term. Next, we isolate $\psi'(u)$, which yields
\begin{equation}
(1 - u^2)\,\psi'(u) 
= - \frac{2 g_{_{\psi\gamma\gamma}} K}{s} b(u)^2 - ct.
\end{equation}

Requiring $\psi \in C^1([-1,1])$, it follows that $\psi$ is bounded in this interval, and therefore
\begin{equation}
ct = - \frac{2 g_{_{\psi\gamma\gamma}} K}{s} b(1)^2 
= - \frac{2 g_{_{\psi\gamma\gamma}} K}{s} b(-1)^2\,.
\end{equation}

And if $g_{_{\psi\gamma\gamma}}$ and $K$ are nonzero, we get
\begin{equation}
b(1)
= b(-1)\,, \quad\text{or}\quad b(1)
=- b(-1)\,.
\end{equation}

Therefore,  {smooth configurations must satisfy one of the two boundary conditions above for $b$. The first possibility, $b(1)=b(-1)$, immediately implies $Q_m=0$ through Eq.~\eqref{magnetic_charge}. The second, $b(1)=-b(-1)$, together with the definite parity of the physical fields, forces the electric charge defined in Eq.~\eqref{electric_attractor} to vanish.}

{Therefore,}  regular axionic rotating attractors are necessarily purely electric or purely magnetic. This is precisely the non‑perturbative version of the perturbative finding around NHEKN/NHEK obtained in Secs. \ref{NHEKN} and \ref{NHEK}. Hence, any rotating dyonic extremal black hole, solution of the field equations \eqref{EME}-\eqref{DE}, cannot have a smooth NHEG limit.
%%%%%%%%%%%%%%%%%%%%%%%%%%%%%%%%%%%%%%%%%%%%%%%%%%%%%%%%%%%%%%%%%%%%%%%%%%%%%%%%%%%%%%%

\bigskip

\section{Bulk black holes}\label{sec4}

We now turn from the near-horizon problem to the bulk description of the gravitational and electromagnetic-axionic fields in a four-dimensional black hole spacetime, in the source-free region outside the horizon. Rotating bulk black hole solutions away from extremality in this model have already been constructed numerically in \cite{Burrage:2023zvk}. Here we compute such configurations again, with the specific goal of studying and characterising the behaviour of the solutions close to extremality and in the extremal limit itself. In this Section, we focus on the physical quantities that will be used to characterise the solutions presented in Sec.~\ref{sec7}, while the explicit metric ansatz, boundary conditions and numerical scheme are collected in Appendix~\ref{sec:numerics}.

In an asymptotically flat, axially symmetric stationary spacetime, the Komar integrals allow for the representation of the total mass and angular momentum through the 2-sphere at spacelike infinity, utilizing the Killing fields denoted by $\xi$ and $\eta$
\begin{equation}\label{mass_komar}
    M=- \dfrac{1}{8\pi}\int_{S^2_{\infty}}\star d\xi=- \dfrac{1}{8\pi}\int_{\mathcal{H}}\star d\xi - \dfrac{1}{4\pi}\int_{\Sigma}\star R(\xi)\,,
\end{equation}
\begin{equation}\label{ang_komar}
    J= \dfrac{1}{16\pi}\int_{S^2_{\infty}}\star d\eta= \dfrac{1}{16\pi}\int_{\mathcal{H}}\star d\eta + \dfrac{1}{8\pi}\int_{\Sigma}\star R(\eta)\, ,
\end{equation}
which yields  the generalized Smarr formula   {\cite{Carter:2009nex,heusler_1996,RASHEED1995379,Kleihaus:2003sh,Compere:2007vx,Ortin:2022uxa,PhysRevLett.30.71,Bardeen:1973gs,Astefanesei:2018vga}}
\begin{equation}\label{mass_formula}
    M=2\Omega_H J + \dfrac{\kappa}{4\pi}A_{\mathcal{H}}+ \Phi_{\mathcal{H}} Q_e + \Psi_{\mathcal{H}} Q_m \, ,
\end{equation}
where
\begin{equation}
\kappa^2=-\frac{1}{2}(\nabla^a \chi^b)\left(\nabla_a \chi_b\right)\Big|_{\mathcal{H}}~,
\end{equation}
with $\chi=\xi+\Omega_H\;\eta$ the horizon generator. The electric and magnetic potentials, $\Phi$ and $\Psi$, are defined through Eq.~\eqref{electr_potentials} with the Killing vector $\chi$ and are constant over the horizon and chosen to vanish at infinity \cite{Prabhu:2015vua,Hajian:2022lgy}. The quantity  $\kappa$ is the surface gravity which determines the 
Hawking temperature $T_{\mathcal{H}}=\kappa/(2\pi)$
and $A_{\mathcal{H}}$
the event horizon area
(with the BH entropy $S=A_{\mathcal{H}}/4$).  Extremal black holes have vanishing Hawking temperature, $T_{\mathcal{H}}=0$, and their
Killing horizons are therefore referred to as degenerate horizons. Despite their zero temperature,
extremal black holes generically have non-vanishing entropy.

We have identified the electric/magnetic charge
\begin{equation}\label{Qe_QM}
    Q_e= -\dfrac{1}{4\pi}\int_{\mathcal{H}}\star\mathcal{F}+g_{_{\psi\gamma\gamma}}\,\psi \mathcal{F}\,,\qquad Q_m=-\dfrac{1}{4\pi}\int_{\mathcal{H}}\mathcal{F}\,.
\end{equation}

Solutions are invariant under shifting the axion field by a constant, $\psi\rightarrow \psi + \theta$, with $\theta$ a real constant. This symmetry implies the existence of a conserved current, $d\, \mathrm{J} =0$, given by \cite{Rakhmanov:1993yd}
\begin{equation}
   \mathrm{J}= \star d \psi-2g_{_{\psi\gamma\gamma}}\mathcal{A}\land\mathcal{F}\, ,
\end{equation}
which remains conserved under gauge transformations.  By asymptotic flatness, the axion field asymptotes as 
\begin{equation}
\psi=\psi_{\infty}-\frac{D}{r}+\mathcal{O}\left(\frac{1}{r^2}\right)\,,
\end{equation}
where  $\psi_{\infty}$ is a constant (assumed to be zero without loss of generality) and $D$ is the scalar monopole
\cite{Prabhu:2018aun,Pacilio:2018kdk}. By   {integrating}  the Noether current, one can show that the axionic hair is of the secondary type   {in the sense that the scalar monopole is not an independent quantity but is fixed by the electromagnetic data \cite{Coleman:1991ku,Prabhu:2018aun,Pacilio:2018kdk,Ballesteros:2023iqb,HerdeiroRadu2015}}
\begin{equation}
\label{scalarcharge}
   D= 4g_{_{\psi\gamma\gamma}}\ \Phi_{\mathcal{H}}\,Q_m\,.
\end{equation}

Therefore, the electromagnetic field   {sources}  the axion; if the former vanishes, the latter trivializes.
  {Moreover}, the scalar   {monopole vanishes for}  purely electric or purely magnetic configurations.

\medskip

Finally,  
let us remark that we display quantities measured in 
terms of ADM mass,
and
introduce the reduced quantities
\begin{eqnarray}
\label{scale1}
j\equiv \frac{J}{M^2}\ ,~~
q\equiv \frac{Q_e}{M}\ , ~~
a_{\mathcal{H}}\equiv \frac{A_{\mathcal{H}}}{16\pi M^2}\ , ~~
t_{\mathcal{H}}\equiv 8\pi T_{\mathcal{H}} M \ ,~~
\omega_H \equiv M \Omega_H~.
\end{eqnarray}
We also define the reduced Ricci and Kretschmann scalars, evaluated at their maximum values
\begin{equation}
    R_{\text{max}}=\text{Max}(R)\, M^2,\qquad K_{\text{max}}=\text{Max}(K)\,M^4\,.
\end{equation}

%%%%%%%%%%%%%%%%%%%%%%%%%%%%%%%%%%%%%%%%%%%%%%%%%%%%%%%%%%%%%%%%%%%%%%%%%%%%%%%%%%%%%%%%

To better understand the singularity structure of the numerical solutions, let us also analyse  curvature scalars of such solutions. For Kerr and Kerr-Newman, the Ricci scalar vanishes everywhere.  However, for all of these solutions, there is a clear singularity $r=0$ when looking at the Kretschmann curvature invariant, $K=R_{\mu \nu \rho \sigma} R^{\mu \nu \rho \sigma}$. Hence, the singularity is hidden by the horizon, and the Kretschmann scalar is smooth everywhere else including on the horizon.

\bigskip

%%%%%%%%%%%%%%%%%%%%%%%%%%%%%%%%%%%%%%%%%%%%%%%%%%%%%%%%%%%%%%%%%%%%%%%%%%%%%%%%%%%%%%%%
\section{Numerical results}\label{sec7}

We now combine the near‑horizon construction with the bulk solutions and chart the space of rotating, extremal black holes in the EMA model. Throughout this section we restrict to the regular, purely electric branch identified by the NHEG analysis. Details on the numerical construction are given in Appendix \ref{sec:numerics}. We present results at representative values of the axion-photon coupling, $g_{_{\psi\gamma\gamma}}=1$ and $g_{_{\psi\gamma\gamma}}=\sqrt{3/2}$, and we compare horizon data extracted from the entropy‑function extremization with the corresponding quantities measured on the extremal black‑hole families.

\begin{figure}[h]

			\centering 
\includegraphics[width=8cm]{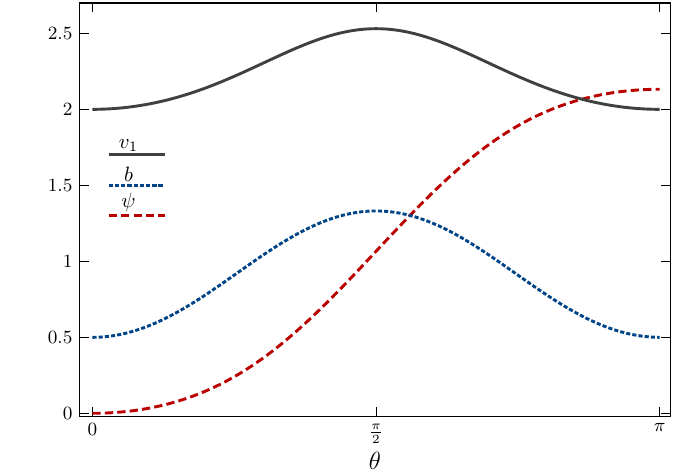}
	\caption{
	{\small
  {Profile functions $v_1(\theta)$, $b(\theta)$ and $\psi(\theta)$}  of a typical rotating near-horizon solution with   {$g_{_{\psi\gamma\gamma}}=1$. The solution is regular on the deformed $S^2$; $v_1$ and $b$ are even in $\theta$, while $\psi$ is odd.}
 }
		\label{profile_attrac}
  }
\end{figure}

Solving the near‑horizon ODE system (\ref{eq:NH-EMA}) in the electric sector yields smooth attractors with $(v_1(\theta),b(\theta),\psi(\theta))$ regular on the deformed $S^2$ and compatible with the $SO(2,1)\times U(1)$ isometry. A typical profile is displayed in Fig.~\ref{profile_attrac} for a given solution with $g_{_{\psi\gamma\gamma}}=1$. Solutions have a definite parity: the metric function $v_1$ and the gauge potential  $b$  are even in $\theta$, while the axion $\psi$ is odd.
%%%%%%%%%%%%%%%%%%%%%%%%%%%%%%%%%%%%%%%%%%%%%%%%%%%%%%
%%%%%%%%%%%%%%%%%%%%%%%%%%%%%%%%%%%%%%%%%%%%%%%%%%%%%%%%

A first lesson is that the near-horizon and bulk pictures agree precisely as long as the attractor exists and both the attractor and the corresponding parent bulk black hole are regular. As established in Sec. \ref{attractors}, smooth rotating axionic attractors occur only for purely electric or purely magnetic charge configurations. This statement holds both perturbatively around NHEK/NHEKN and non‑perturbatively at the level of the full near‑horizon ODE system, and it fixes how we seed the bulk families.  In practice, starting from the purely electric attractors, we   {construct}  global extremal solutions that interpolate between the NHEG and asymptotic flatness; their horizon data match the   {entropy-function}  extremum and thus exhibit standard attractor behaviour, in the sense that  the horizon data   {uniquely determine the bulk}  black hole.

%%%%%%%%%%%%%%%%%%%%%%%%%%%%%%%%%%%%%%%%%%%%%%%%%%%%%%
\begin{figure}[h]
	\makebox[\linewidth][c]{%
		\begin{subfigure}[b]{8cm}
			\centering 
\includegraphics[width=8cm]{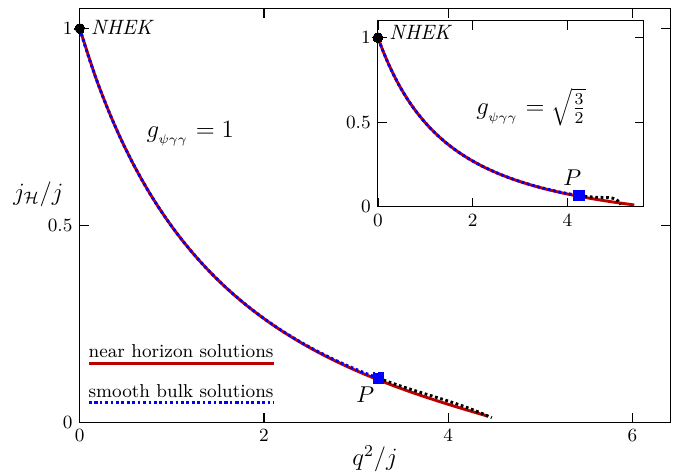}
		\end{subfigure}%
        \hspace{0.5cm}
		\begin{subfigure}[b]{8cm}
			\centering 
\includegraphics[width=8cm]{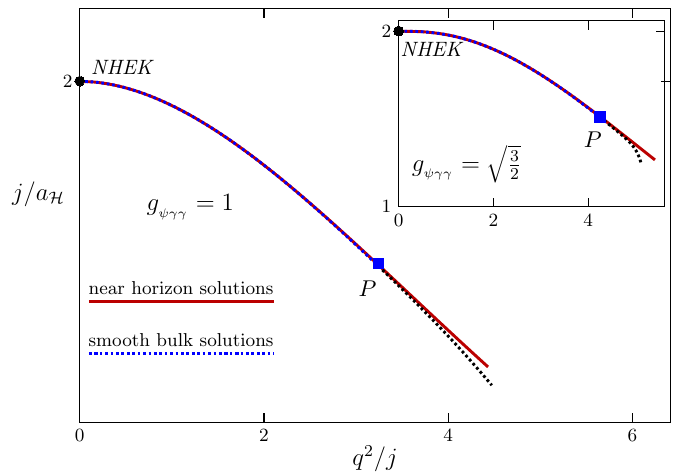} 
		\end{subfigure}%
  } 

	\caption{
	{\small   {(Left Panel) The ratio $j_H/j$ as a function of $q^2/j$. (Right Panel) The ratio $j/a_H$ as a function of $q^2/j$. One can see a}  comparison between the results for 
extremal black hole (dashed) solutions (blue curve)
and near-horizon configurations (red smooth curve)  {for $g_{_{\psi\gamma\gamma}}=1$, while the insets show the corresponding curves for $g_{_{\psi\gamma\gamma}}=\sqrt{3/2}$}. The two   {descriptions}  coincide up to the critical configuration ($P$),  {beyond which}  the bulk   {branch extends}  into  non-smooth configurations (black dotted  curves).
}
		\label{qj}
  }
\end{figure}

%%%%%%%%%%%%%%%%%%%%%%%%%%%%%%%%%%%%%%%%%%%%%%%%%%%%%%%

To quantify the match between horizon and bulk data, we track the ratios ($j_H/j$) and ($j/a_H$) (see Fig.~\ref{qj}). Within numerical accuracy the near‑horizon predictions and the global solutions overlay along the smooth branch and peel away only past a point ($P$). In this first branch, we verified that the reduced Ricci scalar and Kretschmann invariant remain finite all the way to the horizon, in line with the usual behaviour of extremal Kerr/Kerr-Newman. Past the point ($P$), there exists a breakdown in the differentiability of the axionic field with its first derivative presenting a discontinuity; implying, therefore, that the horizon is no longer smooth. For both couplings, the near‑horizon predictions (solid red) overlay the bulk extremal families (blue dashed) along an extended interval. This confirms that, as long as the extremal configurations are smooth, the attractor mechanism fully controls the horizon data and is effectively decoupled from the asymptotics.

 {Since the transition is identified numerically, the precise location of $(P)$ is conventional within numerical resolution; in the plots, we define $(P)$ as the onset of the regime where the attractor and bulk extremal data begin to separate, coinciding, within numerical accuracy, with the loss of smoothness of the bulk axion field through a discontinuity in its first derivative.}

\begin{figure}[!h]
	 \makebox[\linewidth][c]{%
		\begin{subfigure}[b]{8cm}
			\centering 
\includegraphics[width=8cm]{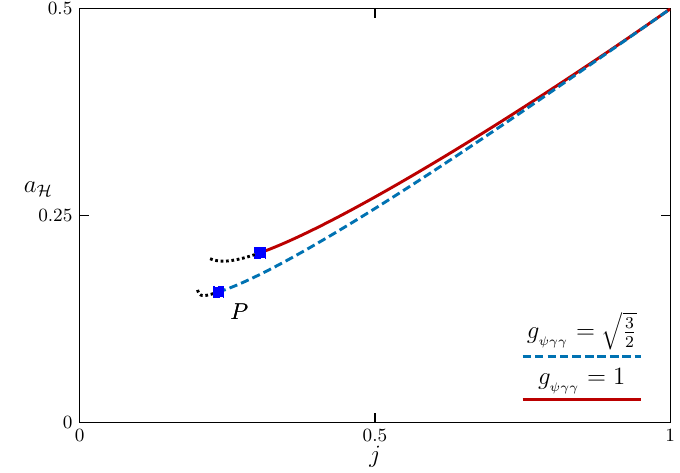}
		\end{subfigure}%
            \hspace{0.5cm}
		\begin{subfigure}[b]{8cm}
			\centering 
\includegraphics[width=8cm]{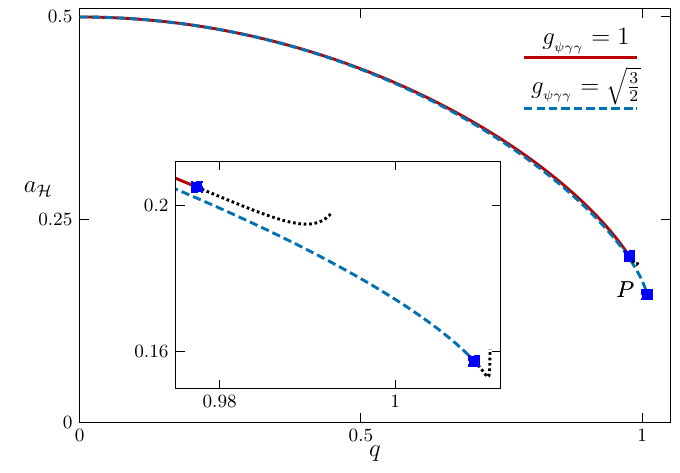} 
		\end{subfigure}%
  } 
	\caption{
	{\small
  {(Left Panel) The reduced horizon area $a_H$ as a function}  of   {the reduced angular momentum $j$. (Right Panel) The reduced horizon area $a_H$ as a function of the reduced electric charge $q$. The solid red and blue dashed curves correspond to}  bulk extremal black hole solutions  {with $g_{_{\psi\gamma\gamma}}=1$ and $g_{_{\psi\gamma\gamma}}=\sqrt{3/2}$, respectively, while the black dotted segments indicate the continuation beyond the critical configuration ($P$). In the inset of the right panel, a zoomed-in visualization of the neighbourhood of ($P$)}.}
		\label{extremal_prop}
  }
\end{figure}

 Starting from NHEK, as we increase the electric charge and simultaneously decrease the spin toward the static regime, the extremal bulk solutions encounter a critical configuration ($P$). Up to ($P$), the near‑horizon and global descriptions coincide. Beyond ($P$), we can continue to construct extremal configurations numerically  (black dotted curves), but they are no longer smooth. Consequently, the NHEG is no longer fixing/describing the horizons of these configurations. In other words, the physical properties of the NHEG and of the extremal black holes start to deviate. The location of ($P$) shifts mildly with ($g_{_{\psi\gamma\gamma}}$), but the structure persists. The plots in Fig. \ref{qj} (see also Fig. \ref{extremal_prop}) illustrate these statements by comparing near‑horizon data (smooth red curves) to the extremal bulk families (blue curves), together with the non‑smooth continuations (black dotted curves) that appear beyond ($P$).

\begin{figure}[h]
	\makebox[\linewidth][c]{%
		\begin{subfigure}[b]{8cm}
			\centering 
\includegraphics[width=8cm]{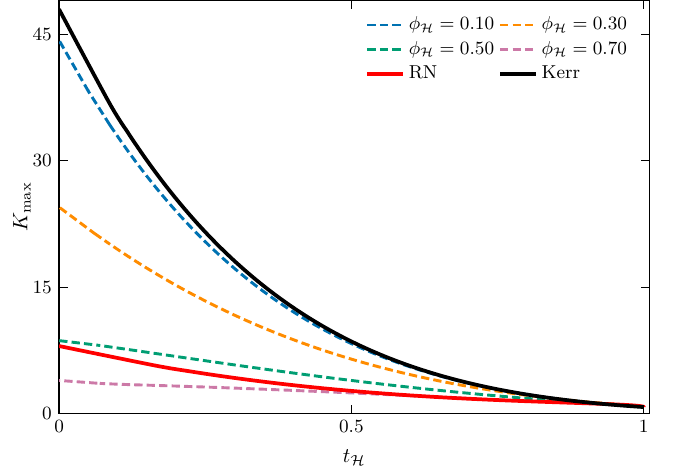}
		\end{subfigure}%
        \hspace{0.5cm}
		\begin{subfigure}[b]{8cm}
			\centering 
\includegraphics[width=8cm]{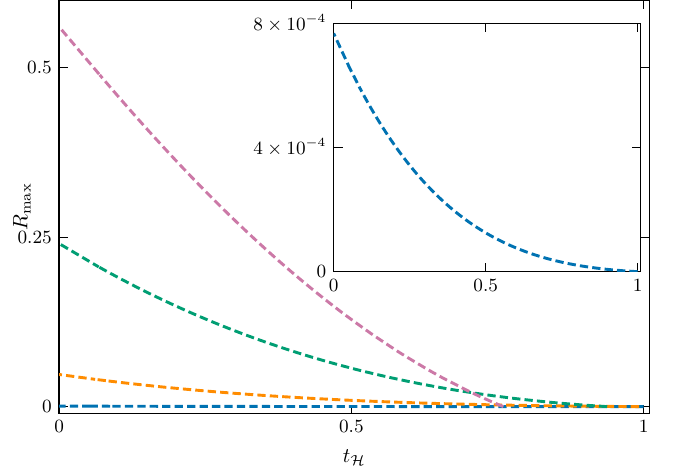} 
		\end{subfigure}%
  } 

	\caption{
	{\small (Left Panel) The reduced Kretschmann scalar curves, with constant electric potential on the horizon, as a function of the temperature. (Right Panel) The reduced Ricci scalar curves, with constant electric potential on the horizon, as a function of the temperature (color code used for the legend is given on the Left Panel). In the inset, a zoomed-in visualization of the curve with constant $\Phi_{\mathcal{H}}=0.1$.}
		\label{scalars}
  }
\end{figure}

Although our primary focus is on extremal solutions, we also construct non‑extremal configurations. As seen in Fig.~\ref{scalars}, the $t_{\mathcal H}$‑dependence of $K_{\mathrm{max}}$ and $R_{\mathrm{max}}$ at fixed $\Phi_{\mathcal H}$ connects smoothly to the extremal values along the regular branch. The reference Kerr and Reissner-Nordström curves included for orientation delimit the expected ranges; our EMA data lie within those envelopes throughout the scans we performed.
 %%%%%%%%%%%%%%%%%%%%%%%%%%%%%%%%%%%%%%%%%%%%%%%%%%%%%%

Taken together, the results exhibit a clear pattern. (i) In the purely electrically charged sector, smooth rotating axionic attractors exist and are realized by asymptotically flat extremal black holes whose horizons are fully controlled by the entropy‑function extremum. (ii) The matched attractor-bulk branch terminates at a critical configuration ($P$), beyond which the extremal solutions persist but lose smoothness (loss of $C^1$ in the axion at the horizon). (iii) These statements hold qualitatively for both values of $g_{_{\psi\gamma\gamma}}$ used here.

 We also followed reduced quantities $(j,\,q,\,a_{\mathcal H})$ along the branch. Figure~\ref{extremal_prop} summarizes representative relations among the reduced area and the spin/charge for extremal solutions. The data vary smoothly up to the critical point and then display the same bifurcation pattern associated with ($P$). 
%%%%%%%%%%%%%%%%%%%%%%%%%%%%%%%%%%%%%%%%%%%%%%%%%%%%%%
%%%%%%%%%%%%%%%%%%%%%%%%%%%%%%%%%%%%%%%%%%%%%%%%%%%
%%%%%%%%%%%%%%%%%%%%%%%%%%%%%%%%%%%%%%%%%%%%%%%%%%%%%%
\begin{figure}[!h]
	\makebox[\linewidth][c]{%
		\begin{subfigure}[b]{8cm}
			\centering
\includegraphics[width=8cm]{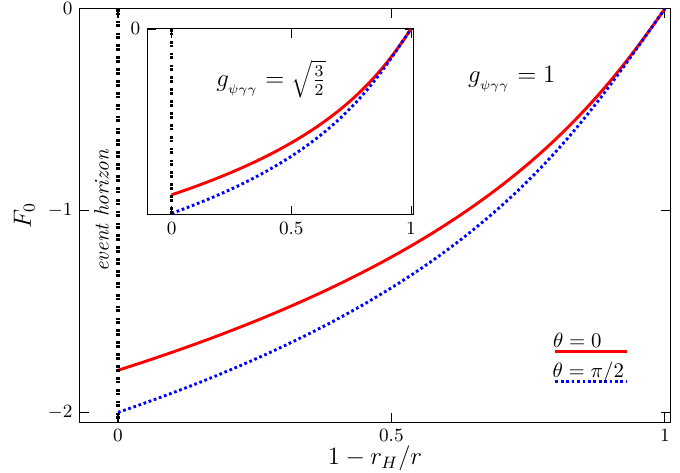}
		\end{subfigure}%
                \hspace{0.5cm}
		\begin{subfigure}[b]{8cm}
			\centering
\includegraphics[width=8cm]{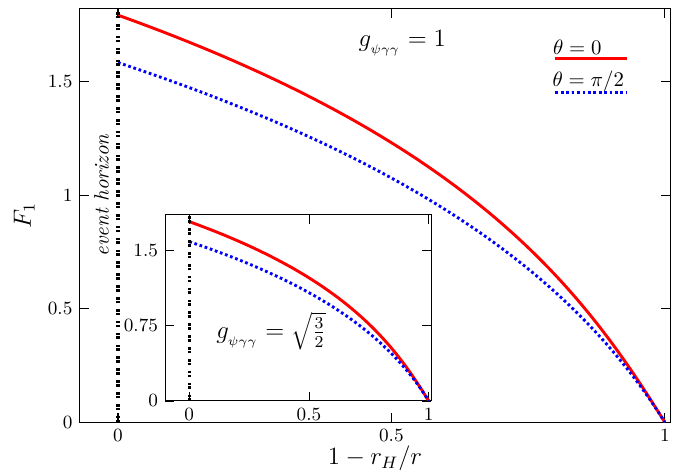}
		\end{subfigure}%
  } 
  \\
  \\
 \makebox[\linewidth][c]{%
		\begin{subfigure}[b]{8cm}
			\centering
\includegraphics[width=8cm]{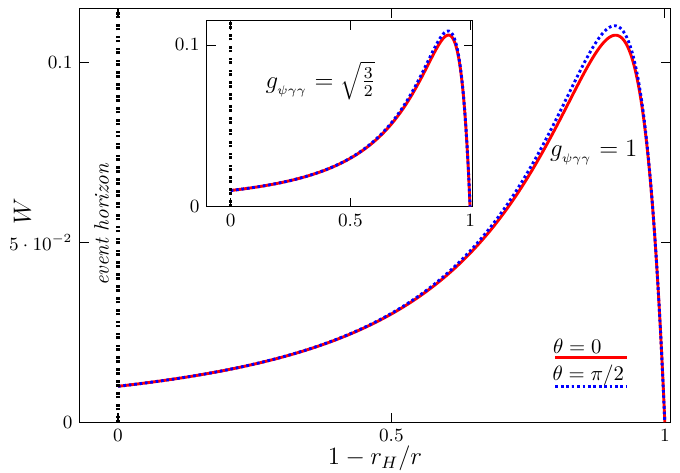}
		\end{subfigure}%
                \hspace{0.5cm}
		\begin{subfigure}[b]{8cm}
			\centering
\includegraphics[width=8cm]{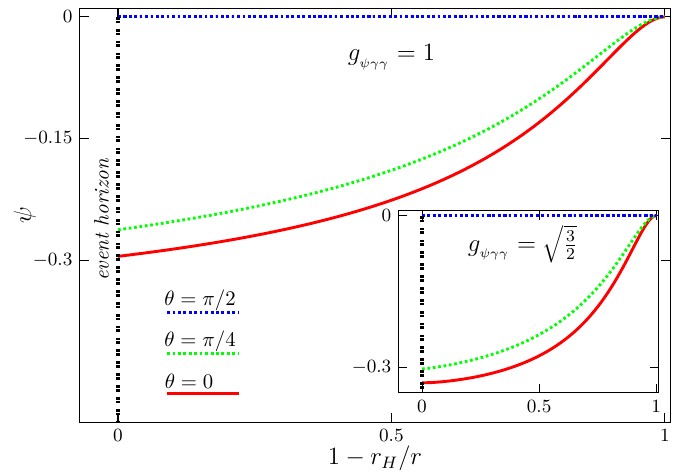}
		\end{subfigure}%
	}
  \\
  \\
 \makebox[\linewidth][c]{%
		\begin{subfigure}[b]{8cm}
			\centering
\includegraphics[width=8cm]{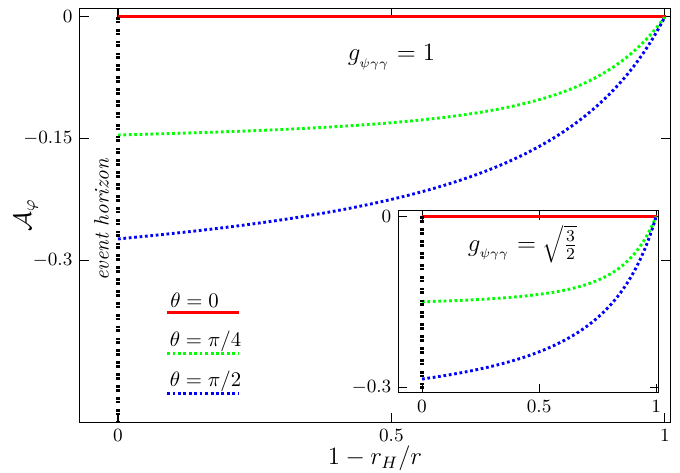}
		\end{subfigure}%
                \hspace{0.5cm}
		\begin{subfigure}[b]{8cm}
			\centering
\includegraphics[width=8cm]{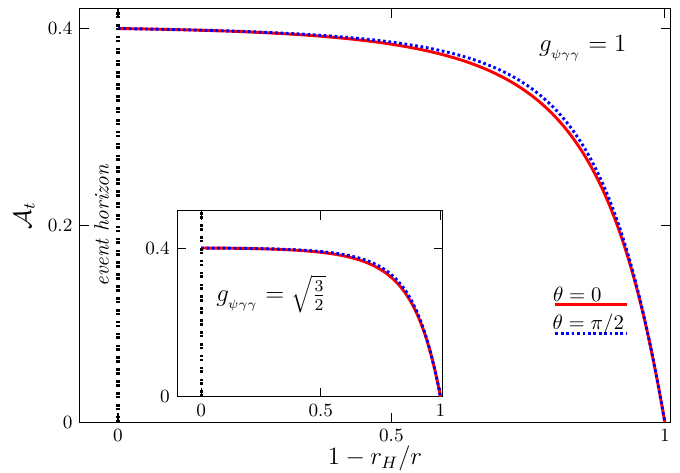}
		\end{subfigure}%
	} 
 \\
	\caption{
	{\small
Profile functions of a typical extremal solution with $g_{_{\psi\gamma\gamma}}=1$, $r_H=0.10$,
$\Omega_H=1.0$,
$\Phi_{\mathcal{H}}=0.40$, 
$vs.$ the compactified radial
coordinate $1-r_H/r$,  
 for several different polar angles $\theta$. The insets show the corresponding functions for a solution with  $g_{_{\psi\gamma\gamma}}=\sqrt{\frac{3}{2}}$ with the same
 input parameters $\{r_H,\Omega_H,\Phi_{\mathcal{H}} \}$.
}
		\label{sol1}
  }
\end{figure}
%%%%%%%%%%%%%%%%%%%%%%%%%%%%%%%%%%%%%%%%%%%%%%%% 

Within each smooth branch we find the same qualitative behaviour for different values of ($g_{_{\psi\gamma\gamma}}$). Horizon quantities vary smoothly with the control parameters and the profiles $(F_0, F_1, W; \psi; \mathcal{A}_\varphi, \mathcal{A}_t)$ exhibit the expected  behaviour (no conical singularities, regular poles, continuous derivatives), and the Komar/Smarr diagnostics from Section 4 remain satisfied along the branch to within our numerical tolerances. The attractor-bulk matched region thus behaves in the standard way familiar from rotating attractors: horizon data are fixed by extremizing the entropy function for an ($SO(2,1)\times U(1)$)‑invariant throat, and the corresponding global solutions exist and interpolate to flat infinity.   

 The typical profiles of $(F_0,F_1,W;\,\psi;\mathcal{A}_\varphi,\mathcal{A}_t)$ are shown in Fig.~\ref{sol1} vs.\ the compactified coordinate $1-r_H/r$ at two polar angles. $F_0$ and $F_1$ exhibit the expected behaviour at the axis and obey $F_1=-F_0$ there (absence of conical singularities up to numerical accuracy). The angular velocity function $W$ is regular, tending to $r_H^2\Omega_H$ at the horizon, and $A_t$ approaches the constant potential $\Phi_{\mathcal H}$. The axion $\psi$ and the magnetic potential $\mathcal{A}_\varphi$ are smooth and axisymmetric across the exterior. The insets show that changing $g_{_{\psi\gamma\gamma}}$ from $1$ to $\sqrt{3/2}$ leaves the qualitative behaviour intact. 

%%%%%%%%%%%%%%%%%%%%%%%%%%%%%%%%%%%%%%%%%%%%%%%%%%%%%%%

\section{Further remarks}

We have analysed rotating attractors and extremal black holes with axionic hair in four dimensions within Einstein-Maxwell-axion theory, combining the entropy formalism with a direct bulk computation of asymptotically flat, axisymmetric solutions. On the near-horizon side, the EMA attractor equations admit regular rotating solutions only in the purely electric or purely magnetic sectors. This restriction was first seen perturbatively around NHEK/NHEKN and then established non‑perturbatively by integrating the axion equation, which enforces regularity at the poles and rules out dyonic data. These results are intrinsic to the NHEG and do not rely on asymptotic information. 

Focusing on the purely electric sector, we constructed one‑parameter families of extremal, asymptotically flat solutions that interpolate smoothly between the attractor and spatial infinity for two representative values of the axion-photon coupling ($g_{_{\psi\gamma\gamma}}$). Along these families, the horizon quantities extracted from the PDE solutions (area, angular momentum, electric charge, and the horizon contribution to $J$) coincide with those predicted by extremizing the entropy function. In particular, the horizon data depend only on the conserved charges $(J,Q_e)$ and are insensitive to asymptotic moduli, as expected from the attractor mechanism.

Increasing the electric charge while decreasing the angular momentum, approaching the static regime, leads to a critical configuration ($P$) beyond which the extremal branch can be continued numerically but is no longer smooth. Concomitantly, the near-horizon and bulk descriptions peel apart—the NHEG constructed from the entropy function no longer captures the horizon data of the extremal configurations past ($P$). The appearance of the critical point ($P$) and the non‑smooth extremal continuations are aligned with a consensus that is being formed in the literature. They emphasize that the near‑horizon analysis is a powerful construction when studying extremal black holes, but does not guarantee the existence—or smoothness—of a corresponding asymptotically flat black hole.

Taken together, these findings organize the extremal sector  {of rotating black holes in the EMA model}  as follows. (i) If the extremal horizon is smooth, then a   {near-horizon}  extremal geometry (rotating attractor) exists  {\cite{Kunduri:2007vf,Kunduri:2013gce}}  and controls the horizon data{, in agreement with the attractor construction \cite{Astefanesei:2006dd,Astefanesei:2007bf,Sen:2005iz}}. (ii) No smooth rotating dyonic attractors arise: the axion equation, together with regularity at the poles, excludes them non-perturbatively, in agreement with the perturbative analysis around NHEK/NHEKN. (iii) Even within the   {purely}  electric sector,   {for which smooth attractors exist, the attractor--bulk}  agreement terminates at a point ($P$); beyond ($P$) we can still construct zero  temperature configurations, but the axion field presents a discontinuity in the derivatives   {at the horizon, and hence those configurations do not admit a smooth NHEG. This should not be understood as saying that such an NHEG cannot exist as a geometry in its own right for those parameter values, but rather that it is not realized as the smooth near-horizon geometry associated with that extremal black hole branch}.

A natural question is whether smooth attractors exist in the full Einstein-Maxwell-dilaton-axion theory. We have therefore attempted to extend our analysis to this more general model. Within a perturbative and numerical NHEG construction at generic dilaton and axion couplings $(\gamma,\,g_{_{\psi\gamma\gamma}})$, we did not find regular rotating solutions; regularity was recovered only at the special couplings $\gamma=g_{_{\psi\gamma\gamma}}=1$, which reproduces the Kerr-Sen attractor.  Two intriguing correlations appear for which we do not yet have a clear explanation. First, for EMDA with generic couplings, the sigma-model target space is symmetric only at $\gamma=g_{_{\psi\gamma\gamma}}=1$, which gives rise to the Kerr-Sen solution; when the axion is absent, it is symmetric at $\gamma=0$ and $\gamma=\sqrt{3}$, leading to the Kerr-Newman and Kaluza-Klein black holes, respectively. The interesting point is that these are also the only couplings that do not impose restrictions on the allowed electromagnetic sectors: purely electric, purely magnetic, and dyonic solutions are all possible.

Second, away from these special coupling values, the existence of smooth extremal black holes seems to require extra constraints in the allowed electromagnetic sector. In the purely dilatonic case ($g_{_{\psi\gamma\gamma}}=0$), smooth extremal black holes (and their NHEG) can be constructed in the dyonic sector,  but only in the fairly restrictive scenario in which the electric charge is equal to the magnetic charge, so that the dilaton monopole (the $r^{-1}$ decay) vanishes \cite{Blazquez-Salcedo:2025cpu,Herdeiro:2025blx}. In the axionic model studied here, smooth extremal black holes (and their NHEG) were found only in the purely electric or purely magnetic sectors, and consequently the axionic monopole charge also vanishes.  Of course, there are solutions, already in the Einstein-Maxwell case, for which the extremal limit is pathological. Likewise, as we have shown here, the vanishing of the axionic monopole charge does not guarantee that the extremal solution has a smooth horizon. Hence, understanding this selectiveness remains an interesting open problem.

%%%%%%%%%%%%%%%%%%%%%%%%%%%%%%%%%%%%%%%%%%%%%%%%%%%%%%%%%%%%%%%%%%%%%%%%%%%%%%%%%%%%%%%%%%%%%%%%%%%%%%
\section*{Acknowledgements}
%%%%%%%%%%%%%%%%%%%%%%%%%%%%%%%%%%%%%%%%%%%%%%%%%%%%%%%%%%%%%%%%%%%%%%%%%%%%%%%%%%%%%%%%%%%%%%%%%%%%%%

I would like to thank C. Herdeiro and E. Radu for their guidance and support throughout this work. I am also grateful to R. Gervalle for his helpful comments on an earlier version of the manuscript. This work is supported by CIDMA under the Portuguese Foundation for Science and Technology (FCT, \url{https://ror.org/00snfqn58}) Multi-Annual Financing Program for R\&D Units, grants UID/4106/2025,  UID/PRR/4106/2025, as well as the projects: Horizon Europe staff exchange (SE) programme HORIZON-MSCA2021-SE-01 Grant No. NewFunFiCO-101086251;  2022.04560.PTDC (\url{https://doi.org/10.54499/2022.04560.PTDC}) and 2024.05617.CERN (\url{https://doi.org/10.54499/2024.05617.CERN}). The author is supported by the FCT grant PRT/BD/153349/2021 (\url{https://doi.org/10.54499/PRT/BD/153349/2021}) under
the IDPASC Doctoral Program.

%%%%%%%%%%%%%%%%%%%%%%%%%%%%%%%%%%%%%%%%%%%%%%%%%%%%%%%%%%%%%%%%%%%%%

\appendix
\section{Numerical scheme}
\label{sec:numerics}
%%%%%%%%%%%%%%%%%%%%%%%%%%%%%%%%%%%%%%%%%%%%%%%%%%%%%%%%%%%%%%%%%%%%%%%%%%%%%%%%

\subsection{Numerical scheme for the rotating attractor equations}
\label{sec:numerics-attractor}

  {Evaluating the EMA Lagrangian density (see the action in Eq. \eqref{action}) on the NHEG ansatz \eqref{metric}--\eqref{matter-nh} gives} 
\begin{align}
\sqrt{-g}\,\mathcal{L}
&= \frac{1}{4\pi}\Biggl\{
-g_{_{\psi\gamma\gamma}}\,\mathcal{B}(\theta)\,\psi(\theta)\,b'(\theta)
+\frac{1}{8\beta\sqrt{v_1(\theta)v_2(\theta)}}\Biggl[
-4v_1(\theta)\,b'(\theta)^2
-v_1(\theta)v_2(\theta)\,\psi'(\theta)^2
\notag\\[0.4em]
&\qquad
+\frac{3v_2(\theta)}{v_1(\theta)}\,v_1'(\theta)^2
-v_1'(\theta)v_2'(\theta)
+\frac{v_1(\theta)}{v_2(\theta)}\,v_2'(\theta)^2
\label{eq:LNH}\\[0.4em]
&\qquad
+\beta^2v_2(\theta)\Bigl[K^2v_2(\theta)+4\mathcal{B}(\theta)^2-4v_1(\theta)\Bigr]
-4v_2(\theta)\,v_1''(\theta)
-2v_1(\theta)\,v_2''(\theta)
\Biggr]\Biggr\},
\notag
\end{align}

{with $\mathcal{B}(\theta) \equiv q + K\,b(\theta)$ as defined  in Sec.~\ref{attractors}. Extremizing}  the entropy function{, \eqref{entropy_function},}  with respect to the profile functions 
\(v_1(\theta)\), \(v_2(\theta)\), \(b(\theta)\), \(\psi(\theta)\), and with
respect to the parameter \(\beta\), yields the coupled ordinary differential
equations
\begin{subequations}
\label{eq:A2}
\setlength{\jot}{3pt}
\begin{align}
v_1''(\theta)
&= \frac{1}{4}\Bigl[
      \beta^{2}\bigl(4   \mathcal{B} (\theta)^{2} + 3 K^{2} v_{2}(\theta)\bigr)
      + \frac{4 v_{1}(\theta) b'(\theta)^{2}}{v_{2}(\theta)}
      + \frac{3 v_{1}'(\theta)^{2}}{v_{1}(\theta)}
      - v_{1}(\theta)\bigl(4\beta^{2} + \psi'(\theta)^{2}\bigr)
    \Bigr], \label{eq:NH-v1}\\[2pt]
v_2''(\theta)
&= \frac{1}{4}\Bigl[
      \frac{\beta^{2} v_{2}(\theta)}{v_{1}(\theta)}
        \bigl(-12   \mathcal{B} (\theta)^{2} - 5 K^{2} v_{2}(\theta)\bigr)
    \Bigr] \notag\\
&\quad
   + \frac{1}{4}\Bigl[
      - 12 b'(\theta)^{2}
      + \frac{v_{2}(\theta) v_{1}'(\theta)^{2}}{v_{1}(\theta)^{2}}
      + \frac{2 v_{2}'(\theta)^{2}}{v_{2}(\theta)}
      + v_{2}(\theta)\bigl(4\beta^{2} - \psi'(\theta)^{2}\bigr)
    \Bigr], \label{eq:NH-v2} \\[2pt]
b''(\theta)
&= - K \beta^{2}\,\frac{v_{2}(\theta)}{v_{1}(\theta)}\,   \mathcal{B} (\theta)
   + \frac{b'(\theta)}{2}
     \Bigl(
        -\frac{v_{1}'(\theta)}{v_{1}(\theta)}
        + \frac{v_{2}'(\theta)}{v_{2}(\theta)}
     \Bigr)   - \beta\, g_{_{\psi\gamma\gamma}}\,   \mathcal{B} (\theta)\,
     \sqrt{\frac{v_{2}(\theta)}{v_{1}(\theta)}}\,\psi'(\theta), \\[2pt]
\psi''(\theta)
&= 4\beta\, g_{_{\psi\gamma\gamma}}\,   \mathcal{B} (\theta)\,
     \frac{b'(\theta)}{\sqrt{v_{1}(\theta)v_{2}(\theta)}}    - \frac{1}{2}
     \Bigl(
        \frac{v_{1}'(\theta)}{v_{1}(\theta)}
        + \frac{v_{2}'(\theta)}{v_{2}(\theta)}
     \Bigr)\psi'(\theta),\\[2pt]
0
&= - v_{1}(\theta)^{2}\Bigl[
        4 b'(\theta)^{2}
        + v_{2}(\theta)\bigl(\psi'(\theta)^{2} - 4\beta^{2}\bigr)
    \Bigr] \notag\\
&\quad
   + v_{1}(\theta)\Bigl\{
       \beta^{2} v_{2}(\theta)\bigl(-4   \mathcal{B} (\theta)^{2} - K^{2} v_{2}(\theta)\bigr)
       + 2 v_{1}'(\theta) v_{2}'(\theta)
     \Bigr\}
   + v_{2}(\theta) v_{1}'(\theta)^{2}\label{eq:NH-constraint},
\end{align}
\end{subequations}
 As we also mentioned in Sec.~\ref{attractors}, the metric functions
\(v_1(\theta)\) and \(v_2(\theta)\) are not independent. 
It is convenient to introduce a new function \(f(\theta)\) through
\begin{equation}
  v_2(\theta) = \frac{f(\theta)}{v_1(\theta)}\,.
\end{equation}
Combining Eqs.~\eqref{eq:NH-v1}, \eqref{eq:NH-v2} and
\eqref{eq:NH-constraint}, we obtain
\begin{equation}
  f(\theta) = s^{2} \cos^{2}\!\big[\beta\,(\theta-\theta_{0})\big] \, ,
\end{equation}
where \(s\) and \(\theta_{0}\) are integration constants. Since the original black hole has $S^2$ topology and physical solutions must be regular on the rotation axis
(\(\theta = 0,\pi\)), i.e. they should not exhibit conical singularities.  
This requirement is encoded in the conditions
\begin{eqnarray}
\label{regular}
 \lim_{\theta \to 0}\frac{g_{\varphi\varphi}}{g_{\theta\theta}}
   = \theta^{2} + \dots,
 \qquad
 \lim_{\theta \to \pi}\frac{g_{\varphi\varphi}}{g_{\theta\theta}}
   = (\pi-\theta)^{2} + \dots,
\end{eqnarray}
which fix
\(\theta_{0} = \pi/2\) and \(\beta = 1\).  Hence the metric functions
satisfy the simple relation
\begin{equation}
  v_2(\theta) = \frac{s^{2}\sin^2(\theta)}{v_1(\theta)}\,,
\end{equation}
with \(s\) an integration constant. From \eqref{regular} one further
finds that this constant coincides with the value of \(v_1\) at the
poles of the two-sphere,
\begin{equation}
\label{new-cond}
  s = v_1(0) = v_1(\pi)\,.
\end{equation}

Hence, after the simplification, the system \eqref{eq:A2} can be cast as in \eqref{eq:NH-EMA}.   {Evaluating the entropy function \eqref{entropy_function} on-shell (i.e. at its extremum)  one finds} 
\begin{equation}
    \mathcal{E}=\pi s
\end{equation}

{so the entropy depends only on the constant $s$ and is independent of any asymptotic modulus. Moreover, extremizing the entropy function \eqref{entropy_function} with respect to
the parameters \(q\) and \(K\) yields, respectively, the electric charge and the
angular momentum:}
\begin{align}
Q_e &=
\frac{1}{2}\int_{0}^{\pi}
\Biggl[
\frac{s\,\sin\theta\,\mathcal{B}(\theta)}{v_1(\theta)}
- g_{_{\psi\gamma\gamma}}\,\psi(\theta)\,b'(\theta)
\Biggr]\,d\theta \, , \label{electric_attractor}
\\[0.5em]
J &=
\frac{1}{2}\int_{0}^{\pi}
\Biggl[
- g_{_{\psi\gamma\gamma}}\,b(\theta)\,\psi(\theta)\,b'(\theta)
+ \frac{s\,b(\theta)\sin\theta\,\mathcal{B}(\theta)}{v_1(\theta)}
+ \frac{K\,s^{3}\sin^{3}\theta}{4\,v_1(\theta)^{2}}
\Biggr]\,d\theta \,.
\end{align}

 We numerically solve the near-horizon (``attractor'') equations for the EMA model, which is a one dimensional ODE system, directly on the polar domain $\theta\in[0,\pi]$\footnote{We also performed computations on the domain $u\in[-1,1]$ and obtained the same results. We choose the $\theta-$domain only for the purpose of later comparison with the bulk solution.}. The solver used in this work was developed by the author.

We cast the system as six first-order ODEs for
\begin{equation}
y(\theta)=\big(b,\,b',\,v_1,\,v_1',\,\psi,\,\psi'\big),
\end{equation}
i.e., three second-order equations for $(b,v_1,\psi)$. The right-hand sides are rational in $\sin\theta,\cos\theta$, the unknowns, and parameters $(g_{_{\psi\gamma\gamma}},\,q,\,K,\,s)$.

Regularity imposes the following expansion around $\theta=0$ (and an equivalent expansion is found for $\theta=\pi$)
\begin{align}
v_1(\theta)&=s+v_{12}\;\theta^2+\mathcal{O}(\theta^3),\\
b(\theta)&=Q_m+u_2\;\theta^2+\mathcal{O}(\theta^3),\\
\psi(\theta)&=\psi_0+\psi_2 \; \theta^2+\mathcal{O}(\theta^3),
\end{align}
with
\begin{equation}
    \psi_2=\frac{2 g_{_{\psi\gamma\gamma}} q Q_{m}}{s}\,,\quad v_{12}=\frac{1}{2} \left(q^2+4 Q_{m}^2-s\right)\,.
\end{equation}
These series provide the shooting data at the axes and encode the regularity conditions.

The boundary-value problem is solved by direct integration from $\theta=\varepsilon$ to $\theta=\pi-\varepsilon$ using a classical fourth-order Runge-Kutta (RK4) method on a uniform $\theta$-grid. We employ a small axis cutoff $\varepsilon=10^{-10}$ and up to $3\times10^5$ steps, with the step size halved $(h\rightarrow h/2)$ for the final $\theta$-interval.

Because regularity plus global constraints over-determine the boundary data, we use a nested shooting strategy:
\begin{itemize}
  \item \textbf{Inner shoot (axis data):} bisect on $u_2$ in the north-pole expansion until the south-pole regularity condition $\psi'(\pi-\varepsilon)=0$  is met.
  \item \textbf{Outer shoot (rotation):} bisect on the near-horizon rotation parameter $K$ until the metric normalization matches across the sphere, quantified by
  \[
  \delta \equiv 1 - \frac{v_1(0)}{v_1(\pi-\varepsilon)}=0.
  \]
\end{itemize}

The typical converged values are $|\delta|\lesssim 10^{-14}$ and $|\psi'(\pi-\varepsilon)|\lesssim10^{-7}$. In this setup, we fix $(g_{_{\psi\gamma\gamma}},s,q)$ and vary $Q_m$. We also performed computations shooting the parameter $q$.

In addition to the two shooting residuals, we continuously monitor the constraint equation, which should vanish identically for exact solutions. Numerically, we evaluate its discrete residual on the grid and sum it over all grid points. For all runs presented here, this global measure of the constraint violation remains below $10^{-9}$.

As numerical benchmarks, we have also reproduced the known attractors of Kerr-Newman and Kerr-Sen (see \cite{Astefanesei:2006dd} for the analytical functions). The solver recovers these exact families within the same tolerance: the summed constraint residual and the errors in the physical quantities are again smaller than $10^{-9}$.

%%%%%%%%%%%%%%%%%%%%%%%%%%%%%%%%%%%%%%%%%%%%%%%%%%%%%%%%%%%%%%%%%%%%%%%%%%%%%%%%%%%%%%%%%%%%%%%%
\subsection{Numerical scheme for the rotating black holes}

The metric ansatz is constructed to accommodate the presence of a horizon. The line element is:
\begin{equation}\label{line_BH}
d s^2=-e^{2 F_0} N d t^2+e^{2 F_1}\left(\frac{d r^2}{N}+r^2 d \theta^2\right)+e^{-2 F_0} r^2 \sin ^2 \theta\left(d \varphi-\dfrac{W}{r^2} d t\right)^2\,,
\end{equation}
where
\begin{equation}
N \equiv 1-\dfrac{r_H}{r}\,.
\end{equation}
	and $(F_i, W)$ are functions of the spheroidal coordinates $(r, \theta)$. The gauge and scalar fields are parametrized by
\begin{equation}
\mathcal{A}_\mu d x^\mu=\left(\mathcal{A}_t- \mathcal{A}_{\varphi} \sin\theta\dfrac{W}{r^2}\right)d t+\mathcal{A}_{\varphi}\sin\theta d \varphi\,,\qquad\qquad \psi=\psi(r,\theta)\,.
\end{equation}

If it exists, the black hole solution arising from \eqref{line_BH} will have a nonzero temperature. In order to construct an extremal black hole (zero temperature), using geometric/regularity arguments \cite{CarneirodaCunha:2010cyg}, one sees that the metric might be written as
\begin{equation}\label{line_BH_extre}
d s^2=-e^{2 F_0} N^2 d t^2+e^{2 F_1}\left(\frac{d r^2}{N^2}+r^2 d \theta^2\right)+e^{-2 F_0} r^2 \sin ^2 \theta\left(d \varphi-\dfrac{W}{r^2} d t\right)^2\,.
\end{equation}
\medskip
Finding EMA solutions with the above ansatz requires defining boundary behaviours. We have made the following choices.
	\begin{enumerate}		
		\item at infinity,
		\begin{equation}
			F_i=W=\mathcal{A}_t=\mathcal{A}_\varphi= \psi=0 \, .
		\end{equation}

		\item on the symmetry axis,
		\begin{equation}
			\partial_\theta F_i=\partial_\theta W=\partial_\theta \mathcal{A}_t=\mathcal{A}_\varphi=\partial_\theta\psi=0\,.
		\end{equation}

	\end{enumerate}

 For the metric ansatz~\eqref{line_BH} (or \eqref{line_BH_extre}), the event horizon is located at a surface with constant radial variable, $r=r_H>0$.
The horizon boundary conditions and the numerical treatment of the problem are simplified by introducing a new radial coordinate
\begin{equation}
x=\sqrt{r^2-r_H^2} ~\, ,
\label{x}
\end{equation}
such that the boundary conditions we impose at the horizon are
		\begin{equation}
			\partial_x F_i=\partial_x \mathcal{A}_\varphi=\partial_x\psi=0 \, ,\qquad   W=r_H^2 \Omega_H\,,\qquad \mathcal{A}_t=\Phi_{\mathcal{H}}\, .
		\end{equation}

These conditions are consistent with a near-horizon solution of the form
\begin{eqnarray}
\label{rh}
{\cal F}_i(r,\theta)= {\cal F}_{i0}(\theta)+x^2 {\cal F}_{i2}(\theta)+\mathcal{O}(x^4)\ ,
\end{eqnarray}  
with ${\cal F}_i =\{F_0, F_1,  W; \psi;\mathcal{A}_\varphi,\mathcal{A}_t\}$,
where the essential functions are
${\cal F}_{i0}$. We mention that $(F_0 -F_1)\big |_{r_H}=const.$,
as imposed by a constraint equation and physically related to the constancy of the temperature on the horizon. Moreover, the absence of conical singularities implies also that $
F_1=-F_0 
$
on the symmetry axis.

\medskip

To compute the solutions, we use the finite-difference boundary-value solver CADSOL \cite{SCHONAUER1989279,SCHONAUER1990279,SCHONAUER2001473} (see \cite{Delgado:2022pwo,Herdeiro:2025blx,Herdeiro:2024pmv} for representative applications and implementation details). We discretize the equations on a rectangular grid with ($N_X\times N_\theta$) points and compactify the radial coordinate via $X=\frac{x}{1+x}$, with $x=\sqrt{r^{2}-r_H^{2}}$. This maps the semi-infinite interval ($[0,\infty)$) to ($[0,1]$). Under this change of variables, derivatives transform as
	\begin{equation}
		\mathcal{F}_{, x} \longrightarrow (1-  X)^2 \mathcal{F}_{, X}, \quad \mathcal{F}_{, x x} \longrightarrow (1-X)^4 \mathcal{F}_{, XX}-2(1-X)^3 \mathcal{F}_{, X} \ .
	\end{equation}

We employ an equidistant grid with ($N_X=300$) points covering ($0\le X\le 1$) and ($N_\theta=100$) points covering ($0\le \theta\le \pi$). Also, we do not impose reflection symmetry across the equatorial plane ($\theta=\pi/2$). Nevertheless, the converged numerical solutions exhibit parity symmetry.\footnote{As a consistency check, we solved a subset of configurations on ($0\le\theta\le\pi/2$) and on ($0\le\theta\le\pi$) and verified that the results coincide.}

	%%%%%%%%%%%%%%%%%%%%%%%%%%%%%%%%%%%%%%%%%%%%%%%%%%%%%%%%%%%%%%%%%%%%%%%%%%%%%%
	  \bibliographystyle{hhieeetr}
	\bibliography{biblio}

@book{heusler_1996, place={Cambridge}, series={Cambridge Lecture Notes in Physics}, title={Black Hole Uniqueness Theorems}, DOI={10.1017/CBO9780511661396}, publisher={Cambridge University Press}, author={Heusler, Markus}, year={1996}, collection={Cambridge Lecture Notes in Physics}}

@article{Carter:1970ea,
    author = "Carter, Brandon",
    title = "{The commutation property of a stationary, axisymmetric system}",
    doi = "10.1007/BF01647092",
    journal = "Commun. Math. Phys.",
    volume = "17",
    pages = "233--238",
    year = "1970"
}

@article{Kundt:1966zz,
    author = "Kundt, Wolfgang and Trumper, M.",
    title = "{Orthogonal decomposition of axi-symmetric stationary spacetimes}",
    doi = "10.1007/BF01325677",
    journal = "Z. Phys.",
    volume = "192",
    pages = "419--422",
    year = "1966"
}

@article{Carter:1969zz,
    author = "Carter, Brandon",
    title = "{Killing horizons and orthogonally transitive groups in space-time}",
    doi = "10.1063/1.1664763",
    journal = "J. Math. Phys.",
    volume = "10",
    pages = "70--81",
    year = "1969"
}

@article{Carter:2009nex,
    author = "Carter, Brandon",
    title = "{Republication of: Black hole equilibrium states}",
    doi = "10.1007/s10714-009-0888-5",
    journal = "Gen. Rel. Grav.",
    volume = "41",
    number = "12",
    pages = "2873--2938",
    year = "2009"
}

@article{Yazadjiev:2010bj,
    author = "Yazadjiev, Stoytcho S.",
    title = "{A Classification (uniqueness) theorem for rotating black holes in 4D Einstein-Maxwell-dilaton theory}",
    eprint = "1009.2442",
    archivePrefix = "arXiv",
    primaryClass = "hep-th",
    doi = "10.1103/PhysRevD.82.124050",
    journal = "Phys. Rev. D",
    volume = "82",
    pages = "124050",
    year = "2010"
}

@article{GIBBONS1988741,
title = {Black holes and membranes in higher-dimensional theories with dilaton fields},
journal = {Nuclear Physics B},
volume = {298},
number = {4},
pages = {741-775},
year = {1988},
issn = {0550-3213},
doi = {https://doi.org/10.1016/0550-3213(88)90006-5},
url = {https://www.sciencedirect.com/science/article/pii/0550321388900065},
author = {G.W. Gibbons and Kei-ichi Maeda}
}

@article{Garfinkle:1990qj,
    author = "Garfinkle, David and Horowitz, Gary T. and Strominger, Andrew",
    title = "{Charged black holes in string theory}",
    reportNumber = "UCSB-TH-90-66",
    doi = "10.1103/PhysRevD.43.3140",
    journal = "Phys. Rev. D",
    volume = "43",
    pages = "3140",
    year = "1991",
    note = "[Erratum: Phys.Rev.D 45, 3888 (1992)]"
}

@article{RASHEED1995379,
title = {The rotating dyonic black holes of Kaluza-Klein theory},
journal = {Nuclear Physics B},
volume = {454},
number = {1},
pages = {379-401},
year = {1995},
issn = {0550-3213},
doi = {https://doi.org/10.1016/0550-3213(95)00396-A},
url = {https://www.sciencedirect.com/science/article/pii/055032139500396A},
author = {Dean Rasheed},
}

@article{Kleihaus:2003sh,
    author = "Kleihaus, Burkhard and Kunz, Jutta and Navarro-Lerida, Francisco",
    title = "{Rotating dilaton black holes with hair}",
    eprint = "gr-qc/0306058",
    archivePrefix = "arXiv",
    doi = "10.1103/PhysRevD.69.064028",
    journal = "Phys. Rev. D",
    volume = "69",
    pages = "064028",
    year = "2004"
}

@article{PhysRevLett.30.71,
  title = {Mass Formula for Kerr Black Holes},
  author = {Smarr, Larry},
  journal = {Phys. Rev. Lett.},
  volume = {30},
  issue = {2},
  pages = {71--73},
  numpages = {0},
  year = {1973},
  month = {Jan},
  publisher = {American Physical Society},
  doi = {10.1103/PhysRevLett.30.71},
  url = {https://link.aps.org/doi/10.1103/PhysRevLett.30.71}
}

@phdthesis{Delgado:2022pwo,
    author = "Delgado, Jorge F. M.",
    title = "{Spinning Black Holes with Scalar Hair and Horizonless Compact Objects within and beyond General Relativity}",
    eprint = "2204.02419",
    archivePrefix = "arXiv",
    primaryClass = "gr-qc",
    school = "Aveiro U.",
    year = "2022"
}

@article{Ortin:2022uxa,
    author = "Ortin, Tomas and Pere\~niguez, David",
    title = "{Magnetic charges and Wald entropy}",
    eprint = "2207.12008",
    archivePrefix = "arXiv",
    primaryClass = "hep-th",
    reportNumber = "IFT-UAM/CSIC-22-40",
    doi = "10.1007/JHEP11(2022)081",
    journal = "JHEP",
    volume = "11",
    pages = "081",
    year = "2022"
}

@article{Hajian:2022lgy,
    author = "Hajian, Kamal and Sheikh-Jabbari, M. M. and Tekin, Bayram",
    title = "{Gauge invariant derivation of zeroth and first laws of black hole thermodynamics}",
    eprint = "2209.00563",
    archivePrefix = "arXiv",
    primaryClass = "hep-th",
    doi = "10.1103/PhysRevD.106.104030",
    journal = "Phys. Rev. D",
    volume = "106",
    number = "10",
    pages = "104030",
    year = "2022"
}

@article{Prabhu:2015vua,
    author = "Prabhu, Kartik",
    title = "{The First Law of Black Hole Mechanics for Fields with Internal Gauge Freedom}",
    eprint = "1511.00388",
    archivePrefix = "arXiv",
    primaryClass = "gr-qc",
    doi = "10.1088/1361-6382/aa536b",
    journal = "Class. Quant. Grav.",
    volume = "34",
    number = "3",
    pages = "035011",
    year = "2017"
}

@article{SCHONAUER1989279,
title = {Efficient vectorizable PDE solvers},
journal = {Journal of Computational and Applied Mathematics},
volume = {27},
number = {1},
pages = {279-297},
year = {1989},
note = {Special Issue on Parallel Algorithms for Numerical Linear Algebra},
issn = {0377-0427},
doi = {https://doi.org/10.1016/0377-0427(89)90371-3},
url = {https://www.sciencedirect.com/science/article/pii/0377042789903713},
author = {W. Schönauer and R. Wei\ss}
}

@incollection{SCHONAUER1990279,
title = {Efficient vectorizable PDE solvers},
editor = {Henk A. {van der Vorst} and Paul {van Dooren}},
series = {Advances in Parallel Computing},
publisher = {North-Holland},
volume = {1},
pages = {279-297},
year = {1990},
booktitle = {Parallel Algorithms for Numerical Linear Algebra},
issn = {0927-5452},
doi = {https://doi.org/10.1016/B978-0-444-88621-7.50019-0},
url = {https://www.sciencedirect.com/science/article/pii/B9780444886217500190},
author = {W. SCHÖNAUER and R. WEIß}
}

@article{SCHONAUER2001473,
title = {How We solve PDEs},
journal = {Journal of Computational and Applied Mathematics},
volume = {131},
number = {1},
pages = {473-492},
year = {2001},
issn = {0377-0427},
doi = {https://doi.org/10.1016/S0377-0427(00)00255-7},
url = {https://www.sciencedirect.com/science/article/pii/S0377042700002557},
author = {Willi Schönauer and Torsten Adolph}
}

@article{Kaluza:1921tu,
    author = "Kaluza, Th.",
    title = {{Zum Unit\"atsproblem der Physik}},
    eprint = "1803.08616",
    archivePrefix = "arXiv",
    primaryClass = "physics.hist-ph",
    reportNumber = "HUPD-8401",
    doi = "10.1142/S0218271818700017",
    journal = "Sitzungsber. Preuss. Akad. Wiss. Berlin (Math. Phys. )",
    volume = "1921",
    pages = "966--972",
    year = "1921"
}

@article{HerdeiroRadu2015,
  author       = {Herdeiro, Carlos A. R. and Radu, Eugen},
  title        = {Asymptotically flat black holes with scalar hair: a review},
  journal      = {Int. J. Mod. Phys. D},
  volume       = {24},
  number       = {09},
  pages        = {1542014},
  year         = {2015},
  doi          = {10.1142/S0218271815420146},
  eprint       = {1504.08209},
  archivePrefix= {arXiv},
  primaryClass = {gr-qc}
}

@article{Klein1926,
  author  = {Oskar Klein},
  title   = {Quantentheorie und f{\"u}nfdimensionale Relativit{\"a}tstheorie},
  journal = {Zeitschrift f{\"u}r Physik},
  volume  = {37},
  number  = {12},
  pages   = {895--906},
  year    = {1926},
  date    = {1926-12-01},
  issn    = {0044-3328},
  doi     = {10.1007/BF01397481},
  url     = {https://doi.org/10.1007/BF01397481}
}

@book{Ortin:2015hya,
    author = "Ortin, Tomas",
    title = "{Gravity and Strings}",
    edition = "2nd ed.",
    doi = "10.1017/CBO9781139019750",
    isbn = "978-0-521-76813-9, 978-0-521-76813-9, 978-1-316-23579-9",
    publisher = "Cambridge University Press",
    series = "Cambridge Monographs on Mathematical Physics",
    month = "7",
    year = "2015"
}

@phdthesis{Pacilio:2018kdk,
    author = "Pacilio, Costantino",
    title = "{Black holes beyond general relativity: theoretical and phenomenological developments}",
    school = "SISSA, Trieste",
    year = "2018",
    url={https://iris.sissa.it/handle/20.500.11767/82334#.XGrMFoVS_eQ}
}

@article{Compere:2007vx,
    author = "Compere, Geoffrey",
    title = "{Note on the First Law with p-form potentials}",
    eprint = "hep-th/0703004",
    archivePrefix = "arXiv",
    reportNumber = "ULB-TH-07-10",
    doi = "10.1103/PhysRevD.75.124020",
    journal = "Phys. Rev. D",
    volume = "75",
    pages = "124020",
    year = "2007"
}

@article{Galtsov:1995zm,
    author = "Galtsov, D. V. and Garcia, A. A. and Kechkin, O. V.",
    title = "{Symmetries of the stationary Einstein-Maxwell dilaton - axion theory}",
    doi = "10.1063/1.531212",
    journal = "J. Math. Phys.",
    volume = "36",
    pages = "5023--5041",
    year = "1995"
}

@article{Wells:1998gc,
    author = "Wells, Clive G.",
    title = "{Extending the black hole uniqueness theorems. 2. Superstring black holes}",
    eprint = "gr-qc/9808045",
    archivePrefix = "arXiv",
    reportNumber = "DAMTP-98-106, DAMTP-1998-106",
    month = "8",
    year = "1998"
}

@article{Bokulic:2023oxw,
    author = "Bokuli\'c, Ana and Smoli\'c, Ivica",
    title = "{Generalizations and challenges for the spacetime block-diagonalization}",
    eprint = "2303.00764",
    archivePrefix = "arXiv",
    primaryClass = "gr-qc",
    reportNumber = "ZTF-EP-23-01",
    doi = "10.1088/1361-6382/ace589",
    journal = "Class. Quant. Grav.",
    volume = "40",
    number = "16",
    pages = "165010",
    year = "2023",
    note = "[Erratum: Class.Quant.Grav. 41, 029501 (2024)]"
}

@article{Rakhmanov:1993yd,
    author = "Rakhmanov, Malik",
    title = "{Dilaton black holes with electric charge}",
    eprint = "hep-th/9310174",
    archivePrefix = "arXiv",
    reportNumber = "CALT-68-1885",
    doi = "10.1103/PhysRevD.50.5155",
    journal = "Phys. Rev. D",
    volume = "50",
    pages = "5155--5163",
    year = "1994"
}

@article{Burrage:2023zvk,
    author = "Burrage, Clare and Fernandes, Pedro G. S. and Brito, Richard and Cardoso, Vitor",
    title = "{Spinning black holes with axion hair}",
    eprint = "2306.03662",
    archivePrefix = "arXiv",
    primaryClass = "gr-qc",
    doi = "10.1088/1361-6382/acf9d6",
    journal = "Class. Quant. Grav.",
    volume = "40",
    number = "20",
    pages = "205021",
    year = "2023"
}

@article{Kunduri:2007vf,
    author = "Kunduri, Hari K. and Lucietti, James and Reall, Harvey S.",
    title = "{Near-horizon symmetries of extremal black holes}",
    eprint = "0705.4214",
    archivePrefix = "arXiv",
    primaryClass = "hep-th",
    reportNumber = "DCPT-07-25",
    doi = "10.1088/0264-9381/24/16/012",
    journal = "Class. Quant. Grav.",
    volume = "24",
    pages = "4169--4190",
    year = "2007"
}

@article{Reall:2002bh,
    author = "Reall, Harvey S.",
    title = "{Higher dimensional black holes and supersymmetry}",
    eprint = "hep-th/0211290",
    archivePrefix = "arXiv",
    reportNumber = "QMUL-PH-02-20",
    doi = "10.1103/PhysRevD.70.089902",
    journal = "Phys. Rev. D",
    volume = "68",
    pages = "024024",
    year = "2003",
    note = "[Erratum: Phys.Rev.D 70, 089902 (2004)]"
}

@article{Kunduri:2013gce,
    author = "Kunduri, Hari K. and Lucietti, James",
    title = "{Classification of near-horizon geometries of extremal black holes}",
    eprint = "1306.2517",
    archivePrefix = "arXiv",
    primaryClass = "hep-th",
    reportNumber = "EMPG-13-09",
    doi = "10.12942/lrr-2013-8",
    journal = "Living Rev. Rel.",
    volume = "16",
    pages = "8",
    year = "2013"
}

@article{Astefanesei:2006dd,
    author = "Astefanesei, Dumitru and Goldstein, Kevin and Jena, Rudra P. and Sen, Ashoke and Trivedi, Sandip P.",
    title = "{Rotating attractors}",
    eprint = "hep-th/0606244",
    archivePrefix = "arXiv",
    reportNumber = "TIFR-TH-06-15, HRI-P-06-06-002",
    doi = "10.1088/1126-6708/2006/10/058",
    journal = "JHEP",
    volume = "10",
    pages = "058",
    year = "2006"
}

@article{Astefanesei:2007bf,
    author = "Astefanesei, Dumitru and Yavartanoo, Hossein",
    title = "{Stationary black holes and attractor mechanism}",
    eprint = "0706.1847",
    archivePrefix = "arXiv",
    primaryClass = "hep-th",
    reportNumber = "KIAS-P07013",
    doi = "10.1016/j.nuclphysb.2007.10.015",
    journal = "Nucl. Phys. B",
    volume = "794",
    pages = "13--27",
    year = "2008"
}

@article{Herdeiro:2025blx,
    author = "Herdeiro, Carlos A. R. and Radu, Eugen and dos Santos Costa Filho, Etevaldo",
    title = "{Charged, rotating black holes in Einstein-Maxwell-dilaton theory}",
    eprint = "2506.15798",
    archivePrefix = "arXiv",
    primaryClass = "gr-qc",
    doi = "10.1088/1475-7516/2026/04/005",
    journal = "JCAP",
    volume = "04",
    pages = "005",
    year = "2026"
}

@article{Herdeiro:2024pmv,
    author = "Herdeiro, Carlos and Radu, Eugen and dos Santos Costa Filho, Etevaldo",
    title = "{Spinning Proca-Higgs balls, stars and hairy black holes}",
    eprint = "2406.03552",
    archivePrefix = "arXiv",
    primaryClass = "gr-qc",
    doi = "10.1088/1475-7516/2024/07/081",
    journal = "JCAP",
    volume = "07",
    pages = "081",
    year = "2024"
}

@book{Straumann:2013spu,
    author = "Straumann, Norbert",
    title = "{General Relativity}",
    doi = "10.1007/978-94-007-5410-2",
    publisher = "Springer",
    address = "Dordrecht",
    series = "Graduate Texts in Physics",
    year = "2013"
}

@article{CarneirodaCunha:2010cyg,
    author = "Carneiro da Cunha, Bruno and de Queiroz, Amilcar R.",
    title = "{Kerr-CFT From Black-Hole Thermodynamics}",
    eprint = "1006.0510",
    archivePrefix = "arXiv",
    primaryClass = "hep-th",
    doi = "10.1007/JHEP08(2010)076",
    journal = "JHEP",
    volume = "08",
    pages = "076",
    year = "2010"
}

@article{Horowitz:2023xyl,
    author = "Horowitz, Gary T. and Kolanowski, Maciej and Remmen, Grant N. and Santos, Jorge E.",
    title = "{Extremal Kerr Black Holes as Amplifiers of New Physics}",
    eprint = "2303.07358",
    archivePrefix = "arXiv",
    primaryClass = "hep-th",
    doi = "10.1103/PhysRevLett.131.091402",
    journal = "Phys. Rev. Lett.",
    volume = "131",
    number = "9",
    pages = "091402",
    year = "2023"
}

@article{Compere:2012jk,
    author = "Comp{\`e}re, Geoffrey",
    title = "{The Kerr/CFT Correspondence and its Extensions}",
    eprint = "1203.3561",
    archivePrefix = "arXiv",
    primaryClass = "hep-th",
    doi = "10.12942/lrr-2012-11",
    journal = "Living Rev. Rel.",
    volume = "15",
    number = "1",
    pages = "11--81",
    year = "2012"
}

@article{Sen:2007qy,
    author = "Sen, Ashoke",
    title = "{Black Hole Entropy Function, Attractors and Precision Counting of Microstates}",
    eprint = "0708.1270",
    archivePrefix = "arXiv",
    primaryClass = "hep-th",
    doi = "10.1007/s10714-008-0626-4",
    journal = "Gen. Rel. Grav.",
    volume = "40",
    pages = "2249--2431",
    year = "2008"
}

@article{Horowitz:2022mly,
    author = "Horowitz, Gary T. and Kolanowski, Maciej and Santos, Jorge E.",
    title = "{Almost all extremal black holes in AdS are singular}",
    eprint = "2210.02473",
    archivePrefix = "arXiv",
    primaryClass = "hep-th",
    doi = "10.1007/JHEP01(2023)162",
    journal = "JHEP",
    volume = "01",
    pages = "162",
    year = "2023"
}

@article{Horowitz:2024kcx,
    author = "Horowitz, Gary T. and Santos, Jorge E.",
    title = "{Smooth extremal horizons are the exception, not the rule}",
    eprint = "2411.07295",
    archivePrefix = "arXiv",
    primaryClass = "hep-th",
    doi = "10.1007/JHEP02(2025)169",
    journal = "JHEP",
    volume = "02",
    pages = "169",
    year = "2025"
}

@article{Sen:1992ua,
    author = "Sen, Ashoke",
    title = "{Rotating charged black hole solution in heterotic string theory}",
    eprint = "hep-th/9204046",
    archivePrefix = "arXiv",
    reportNumber = "TIFR-TH-92-20",
    doi = "10.1103/PhysRevLett.69.1006",
    journal = "Phys. Rev. Lett.",
    volume = "69",
    pages = "1006--1009",
    year = "1992"
}

@article{Marsh:2015xka,
    author = "Marsh, David J. E.",
    title = "{Axion Cosmology}",
    eprint = "1510.07633",
    archivePrefix = "arXiv",
    primaryClass = "astro-ph.CO",
    reportNumber = "KCL-PH-TH-2015-50",
    doi = "10.1016/j.physrep.2016.06.005",
    journal = "Phys. Rept.",
    volume = "643",
    pages = "1--79",
    year = "2016"
}

@article{Peccei:1977hh,
    author = "Peccei, R. D. and Quinn, Helen R.",
    title = "{CP Conservation in the Presence of Instantons}",
    reportNumber = "ITP-568-STANFORD",
    doi = "10.1103/PhysRevLett.38.1440",
    journal = "Phys. Rev. Lett.",
    volume = "38",
    pages = "1440--1443",
    year = "1977"
}

@article{Bergstrom:2009ib,
    author = "Bergstrom, Lars",
    title = "{Dark Matter Candidates}",
    eprint = "0903.4849",
    archivePrefix = "arXiv",
    primaryClass = "hep-ph",
    doi = "10.1088/1367-2630/11/10/105006",
    journal = "New J. Phys.",
    volume = "11",
    pages = "105006",
    year = "2009"
}

@article{Svrcek:2006yi,
    author = "Svrcek, Peter and Witten, Edward",
    title = "{Axions In String Theory}",
    eprint = "hep-th/0605206",
    archivePrefix = "arXiv",
    reportNumber = "SLAC-PUB-11894",
    doi = "10.1088/1126-6708/2006/06/051",
    journal = "JHEP",
    volume = "06",
    pages = "051",
    year = "2006"
}

@article{Sen:2005iz,
    author = "Sen, Ashoke",
    title = "{Entropy function for heterotic black holes}",
    eprint = "hep-th/0508042",
    archivePrefix = "arXiv",
    doi = "10.1088/1126-6708/2006/03/008",
    journal = "JHEP",
    volume = "03",
    pages = "008",
    year = "2006"
}

@article{Compere:2015mza,
    author = "Comp{\`e}re, G. and Hajian, K. and Seraj, A. and Sheikh-Jabbari, M. M.",
    title = "{Extremal Rotating Black Holes in the Near-Horizon Limit: Phase Space and Symmetry Algebra}",
    eprint = "1503.07861",
    archivePrefix = "arXiv",
    primaryClass = "hep-th",
    doi = "10.1016/j.physletb.2015.08.027",
    journal = "Phys. Lett. B",
    volume = "749",
    pages = "443--447",
    year = "2015"
}

@article{Compere:2015bca,
    author = "Comp{\`e}re, G. and Hajian, K. and Seraj, A. and Sheikh-Jabbari, M. M.",
    title = "{Wiggling Throat of Extremal Black Holes}",
    eprint = "1506.07181",
    archivePrefix = "arXiv",
    primaryClass = "hep-th",
    doi = "10.1007/JHEP10(2015)093",
    journal = "JHEP",
    volume = "10",
    pages = "093",
    year = "2015"
}

@article{Hajian:2013lna,
    author = "Hajian, K. and Seraj, A. and Sheikh-Jabbari, M. M.",
    title = "{NHEG Mechanics: Laws of Near Horizon Extremal Geometry (Thermo)Dynamics}",
    eprint = "1310.3727",
    archivePrefix = "arXiv",
    primaryClass = "hep-th",
    reportNumber = "IPM-P-2013-033",
    doi = "10.1007/JHEP03(2014)014",
    journal = "JHEP",
    volume = "03",
    pages = "014",
    year = "2014"
}

@article{PhysRevLett.40.279,
  title = {Problem of Strong $P$ and $T$ Invariance in the Presence of Instantons},
  author = {Wilczek, F.},
  journal = {Phys. Rev. Lett.},
  volume = {40},
  issue = {5},
  pages = {279--282},
  numpages = {0},
  year = {1978},
  month = {Jan},
  publisher = {American Physical Society},
  doi = {10.1103/PhysRevLett.40.279},
  url = {https://link.aps.org/doi/10.1103/PhysRevLett.40.279}
}

@article{PhysRevLett.40.223,
  title = {A New Light Boson?},
  author = {Weinberg, Steven},
  journal = {Phys. Rev. Lett.},
  volume = {40},
  issue = {4},
  pages = {223--226},
  numpages = {0},
  year = {1978},
  month = {Jan},
  publisher = {American Physical Society},
  doi = {10.1103/PhysRevLett.40.223},
  url = {https://link.aps.org/doi/10.1103/PhysRevLett.40.223}
}

@article{Coleman:1991ku,
    author = "Coleman, Sidney R. and Preskill, John and Wilczek, Frank",
    title = "{Quantum hair on black holes}",
    eprint = "hep-th/9201059",
    archivePrefix = "arXiv",
    reportNumber = "IASSNS-HEP-91-64, CALT-68-1764, HUTP-92-A003",
    doi = "10.1016/0550-3213(92)90008-Y",
    journal = "Nucl. Phys. B",
    volume = "378",
    pages = "175--246",
    year = "1992"
}

@article{Prabhu:2018aun,
    author = "Prabhu, Kartik and Stein, Leo C.",
    title = "{Black hole scalar charge from a topological horizon integral in Einstein-dilaton-Gauss-Bonnet gravity}",
    eprint = "1805.02668",
    archivePrefix = "arXiv",
    primaryClass = "gr-qc",
    doi = "10.1103/PhysRevD.98.021503",
    journal = "Phys. Rev. D",
    volume = "98",
    number = "2",
    pages = "021503",
    year = "2018"
}

@article{Newman:1965my,
    author = "Newman, E T. and Couch, R. and Chinnapared, K. and Exton, A. and Prakash, A. and Torrence, R.",
    title = "{Metric of a Rotating, Charged Mass}",
    doi = "10.1063/1.1704351",
    journal = "J. Math. Phys.",
    volume = "6",
    pages = "918--919",
    year = "1965"
}

@article{Bardeen:1999px,
    author = "Bardeen, James M. and Horowitz, Gary T.",
    title = "{The Extreme Kerr throat geometry: A Vacuum analog of AdS(2) x S**2}",
    eprint = "hep-th/9905099",
    archivePrefix = "arXiv",
    reportNumber = "NSF-ITP-99-29",
    doi = "10.1103/PhysRevD.60.104030",
    journal = "Phys. Rev. D",
    volume = "60",
    pages = "104030",
    year = "1999"
}

@article{Bardeen:1973gs,
    author = "Bardeen, James M. and Carter, B. and Hawking, S. W.",
    title = "{The Four laws of black hole mechanics}",
    doi = "10.1007/BF01645742",
    journal = "Commun. Math. Phys.",
    volume = "31",
    pages = "161--170",
    year = "1973"
}

@article{Ballesteros:2023iqb,
    author = "Ballesteros, Romina and G\'omez-Fayr\'en, Carmen and Ort\'\i{}n, Tom\'as and Zatti, Matteo",
    title = "{On scalar charges and black hole thermodynamics}",
    eprint = "2302.11630",
    archivePrefix = "arXiv",
    primaryClass = "hep-th",
    reportNumber = "IFT-UAM/CSIC-23-018",
    doi = "10.1007/JHEP05(2023)158",
    journal = "JHEP",
    volume = "05",
    pages = "158",
    year = "2023"
}

@article{Lee:1991jw,
    author = "Lee, Ki-Myeong and Weinberg, Erick J.",
    title = "{Charge black holes with scalar hair}",
    reportNumber = "CU-TP-515",
    doi = "10.1103/PhysRevD.44.3159",
    journal = "Phys. Rev. D",
    volume = "44",
    pages = "3159--3163",
    year = "1991"
}

@article{Balakin:2017nbg,
    author = "Balakin, Alexander B. and Zayats, Alexei E.",
    title = "{Einstein{\textendash}Maxwell-axion theory: dyon solution with regular electric field}",
    eprint = "1703.08858",
    archivePrefix = "arXiv",
    primaryClass = "gr-qc",
    doi = "10.1140/epjc/s10052-017-5073-5",
    journal = "Eur. Phys. J. C",
    volume = "77",
    number = "8",
    pages = "519",
    year = "2017"
}

@article{Nakarachinda:2025bvy,
    author = "Nakarachinda, Ratchaphat and Boonserm, Petarpa and De Felice, Antonio and Tsujikawa, Shinji and Wongjun, Pitayuth",
    title = "{Greybody factors of charged black holes with axion hair}",
    eprint = "2506.18241",
    archivePrefix = "arXiv",
    primaryClass = "gr-qc",
    reportNumber = "YITP-25-93, WUCG-25-07",
    doi = "10.1103/5bcc-gc31",
    journal = "Phys. Rev. D",
    volume = "112",
    number = "6",
    pages = "064055",
    year = "2025"
}

@article{Fernandes:2019kmh,
    author = "Fernandes, Pedro G. S. and Herdeiro, Carlos A. R. and Pombo, Alexandre M. and Radu, Eugen and Sanchis-Gual, Nicolas",
    title = "{Charged black holes with axionic-type couplings: Classes of solutions and dynamical scalarization}",
    eprint = "1908.00037",
    archivePrefix = "arXiv",
    primaryClass = "gr-qc",
    doi = "10.1103/PhysRevD.100.084045",
    journal = "Phys. Rev. D",
    volume = "100",
    number = "8",
    pages = "084045",
    year = "2019"
}

@article{Boskovic:2018lkj,
    author = "Boskovic, Mateja and Brito, Richard and Cardoso, Vitor and Ikeda, Taishi and Witek, Helvi",
    title = "{Axionic instabilities and new black hole solutions}",
    eprint = "1811.04945",
    archivePrefix = "arXiv",
    primaryClass = "gr-qc",
    doi = "10.1103/PhysRevD.99.035006",
    journal = "Phys. Rev. D",
    volume = "99",
    number = "3",
    pages = "035006",
    year = "2019"
}

@article{KIczek:2021vlc,
    author = "KIczek, Bartlomiej and Rogatko, Marek",
    title = "{Axion-like dark matter clouds around rotating black holes}",
    journal = {Phys. Rev. D},
    eprint = "2106.01565",
    archivePrefix = "arXiv",
    primaryClass = "hep-th",
    doi = "10.1103/PhysRevD.103.124021",
    month = "6",
    year = "2021"
}

@article{Blazquez-Salcedo:2025cpu,
    author = "Bl{\'a}zquez-Salcedo, Jose Luis and Herdeiro, Carlos and Radu, Eugen and dos Santos Costa Filho, Etevaldo and Uzawa, Kunihito",
    title = "{Spinning extremal dyonic black holes in {\ensuremath{\gamma}} = 1 Einstein-Maxwell-dilaton theory}",
    eprint = "2512.20698",
    archivePrefix = "arXiv",
    primaryClass = "gr-qc",
    doi = "10.1007/JHEP04(2026)078",
    journal = "JHEP",
    volume = "04",
    pages = "078",
    year = "2026"
}

@article{Astefanesei:2018vga,
    author = "Astefanesei, Dumitru and Ballesteros, Romina and Choque, David and Rojas, Ra{\'u}l",
    title = "{Scalar charges and the first law of black hole thermodynamics}",
    eprint = "1803.11317",
    archivePrefix = "arXiv",
    primaryClass = "hep-th",
    doi = "10.1016/j.physletb.2018.05.005",
    journal = "Phys. Lett. B",
    volume = "782",
    pages = "47--54",
    year = "2018"
}
	 %%%%%%%%%%%%%%%%%%%%%%%%%%%%%%%%%%%%%%%%%%%%%%%%%%%%%%%%%%%%%%%%%%%%%%%%%%%%%%

\end{document}